\documentstyle[12pt,aaspp4,flushrt]{article}

\newcommand{\lit}{$^7$Li~}
\newcommand{\nli}{$\log\epsilon(^7{\rm Li})$~}

\newcommand{\msb}{$M_{\odot}$~}
\newcommand{\ms}{$M_{\odot}$}

\begin{document}

\title{Galactic chemical evolution of Lithium: \\
interplay between stellar sources}

\author{Claudia Travaglio}
\affil{1. Max-Planck Institut f\"ur Astronomie, \\ K\"onigsthul 17, D-69117 Heidelberg,
Germany}
\affil{2. Dipartimento di Astronomia e Scienza dello Spazio, \\ Largo E. Fermi 5, I-50125
Firenze, Italy}

\author{Sofia Randich, Daniele Galli}
\affil{3. Osservatorio Astrofisico di Arcetri, \\ Largo E. Fermi 5, I-50125
Firenze, Italy}

\author{John Lattanzio, Lisa M. Elliott}
\affil{4. Department of Mathematics and Statistics, Monash University, \\
Clayton, Victoria, 3168, Australia}

\author{Manuel Forestini}
\affil{5. Laboratoire d'Astrophysique, Observatoire de Grenoble, Universit\'e
Joseph Fourier, \\ BP 53, F-38041 Grenoble Cedex 9, France}

\author{Federico Ferrini}
\affil{6. Dipartimento di Fisica, Sezione di Astronomia, Universit\`a di
Pisa, \\ Piazza Torricelli 2, I-5600 Pisa, Italy}

\begin{abstract}

In this paper we study the evolution of \lit in the Galaxy considering
the contributions of various stellar sources: type II supernovae,
novae, red giant stars, and asymptotic giant branch (AGB) stars.  We
present new results for the production of \lit in AGB stars via the hot
bottom burning process, based on stellar evolutionary models by
Frost~(1997).  In the light of recent observations of dense
circumstellar shells around evolved stars in the Galaxy and in the
Magellanic Clouds, we also consider the impact of a very high mass-loss
rate episode (superwind) before the evolution off the AGB phase on the
$^7$Li enrichment in the interstellar medium. We compare the Galactic
evolution of \lit obtained with these new \lit yields (complemented
with a critical re-analysis of the role of supernovae, novae and giant
stars) with a selected compilation of spectroscopic observations
including halo and disk field stars as well as young stellar clusters.
We conclude that even allowing for the large uncertainties in the
theoretical calculation of mass-loss rates at the end of the AGB phase,
the superwind phase has a significant effect on the \lit enrichment of
the Galaxy.

\end{abstract}


\section{Introduction}

In spite of many attempts, the evolution of \lit in the Galaxy is not
completely understood.  Lithium has the unique property of being
produced by, at least, three different processes: Big Bang
nucleosynthesis, spallation of Galactic cosmic-ray (GCR) particles on
interstellar matter (ISM) nuclei, and stellar nucleosynthesis.  At the
same time, \lit is easily destroyed by proton captures in stellar
interiors at relatively low temperatures ($\sim 2.5\times 10^6$~K).

Observationally, Population I (hereafter Pop.~I) dwarf stars show a
large dispersion in \lit abundance attributed to different amounts of
depletion. On the contrary the abundance of \lit in the atmosphere of
old, warm ($T_{\rm eff}>5700$~K) Population II (hereafter Pop.~II)
dwarf stars is surprisingly uniform:  $^7$Li/H $\simeq 1.1\times
10^{-10}$ (Spite \& Spite~1982), a value $\sim 10$ times lower than the
maximum abundance observed in Pop.~I stars.  Several studies have
subsequently confirmed the existence of the so-called ``Spite plateau''
in Pop.~II stars, although discrepancies on the actual value of the
plateau still exist at the level of $\sim 0.1$~dex (see e.g., Spite \&
Spite 1986; Hobbs \& Duncan 1987; Rebolo, Beckman, \& Molaro~1988;
Thorburn~1994; Spite et al.  1996; Bonifacio \& Molaro 1997)
\footnote{In this paper we adopt the usual notation \nli
$=\log[N(^7{\rm Li})/N({\rm H})] + 12$. The Spite plateau corresponds
to \nli $\simeq 2.1$. }.

There is an ongoing debate about whether the Spite plateau corresponds
to the (almost) undepleted primordial abundance of \lit, or whether the
primordial value is closer to \nli $\simeq 3.1$--3.3 (the value
measured in the youngest stars and solar system meteorites) and has
been significantly but uniformly depleted in Pop.~II stars (see e.g.
Deliyannis \& Ryan 1997; Bonifacio \& Molaro 1997). On one hand, the
former interpretation requires processes that increase the Galactic
\lit abundance by a factor $\sim 10$ in the first few Gyr. On the other
hand {\it standard models} of stellar evolution which incorporate only
convective mixing predict little ($\sim 0.1$~dex) \lit depletion in
Pop.~II stars (e.g., Deliyannis, Demarque \& Kawaler 1990).  Stellar
models including the effects of non-standard processes (like e.g.
diffusion, winds, turbulence induced by rotational instabilities, slow
mixing driven by angular momentum loss, etc.) predict substantial \lit
destruction in Pop.~II stars (see Pinsonneault~1997 for a review).
These non-standard models predict observable features, like a
dispersion in \lit among halo dwarfs, or correlations between \lit and
T$_{\rm eff}$ or [Fe/H], which have not been confirmed by observations,
suggesting that \lit destruction must have been minimal in these stars
(see e.g., Boesgaard \& Steigman~1985, Bonifacio \& Molaro~1997, Ryan,
Norris, \& Beers~1999, Ryan et al.~2000). It is true, however, that the
presence of a few metal--poor stars with \lit abundances somewhat above
the plateau, as well as other observational evidences in Pop.~I stars
(Deliyannis et al.~1998, Deliyannis 2000) support the idea that halo
dwarfs may have suffered a certain amount of \lit depletion.

Indication of a primordial abundance of \lit in excess of the Spite
plateau value is suggested also by ({\em i}\/) recent determinations of
deuterium abundances in Lyman-$\alpha$ clouds (see e.g. Pettini \&
Bowen~2001 and references therein), and ({\em ii}) observations of the
spectrum of fluctuations in the cosmic background radiation (de
Bernardis et. al.~2000, Tegmark \& Zaldarriaga~2000). In both cases the
value of the baryon-to-photon ratio $\eta_{10}$ results in the range
6--7, rather than 4--5 as usually assumed. This implies a cosmological
abundance of \lit in excess of the Spite plateau value by a factor
2--3, still compatible with the observationally allowed range of
primordial $^4$He. However, one should be very careful in interpreting
these indications since possible sources of systematic errors have
still to be completely understood and eliminated from the observational
data. The issue remains therefore open.

Assuming that the Spite plateau corresponds to the
primordial \lit abundance, the difference between the plateau abundance
and the maximum \lit content in Pop.~I stars underscores the need for one
or more sites of \lit production.  

It is well known that
$^9$Be, $^{10,11}$B, and $^{6,7}$Li can be produced via spallation and
fusion reactions between GCR particles and ISM nuclei (Fowler, Reeves,
\& Silk~1970; Meneguzzi, Audouze \& Reeves~1971). Whereas theoretical
calculations of the amounts of $^9$Be, $^{10,11}$B and $^6$Li produced
by GCR (see e.g. Lemoine, Vangioni-Flam, \& Cass\'e~1998; Fields \&
Olive~1999; Ramaty et al.~2000, and references therein) can roughly
reproduce the observed values, the amount of \lit produced by
spallation processes at the time of formation of the Sun is a factor
5--10 lower than the measured meteoritic value (see also Nissen et
al.~1999 and references therein)\footnote{At lower metallicity,
in particular in the range $-2 <$ [Fe/H] $< -0.5$, the predictions of
GCR spallation models indicate a possible overproduction of $^{6,7}$Li
due to $\alpha$-$\alpha$ reactions (Prantzos et al.~1993, Valle et
al.~2001 in preparation) that may suggest the occurence of a depletion
process in halo stars at least in this metallicity range.}. 

Additional sites of \lit production are therefore required to match the
observational data.  For example, D'Antona \& Matteucci~(1991) proposed
that novae and AGB stars may represent the main sources of Galactic
\lit, whereas Matteucci, D'Antona, \& Timmes~(1995) favoured a model
where the main contributors to \lit enrichment are SNII and AGB stars.
Abia, Isern, \& Canal~(1995) considered the production of \lit by
low-mass AGB stars, whereas Romano et al.~(1999) showed that novae are
required in order to reproduce the growth of \lit with metallicity, and
that the contribution from SNII should be lowered by at least a factor
of two. Finally, Romano et al.~(2001), using recent AGB models by
Ventura, D'Antona, \& Mazzitelli~(2000), concluded that AGB stars are
not substantial \lit producers in the Galaxy.

In this paper we reanalyze the problem afresh with the help of a
numerical model of Galactic chemical evolution and a set of new
models of AGB stars of various masses and metallicities.  We consider
various stellar sources of \lit:  novae, type II supernovae, low-- and
intermediate--mass stars in the red giant branch (RGB) and AGB phase.
All these sources can produce \lit, as shown by theoretical models,
and, in some cases, also confirmed by spectroscopic data.  However,
individual \lit yields are rather uncertain and strongly dependent on
model parameters and assumptions.

The paper is organized as follows. In \S~2 we discuss our sample of
observational data for field stars and open clusters.  In \S~3 we
briefly describe the adopted model for  Galactic chemical evolution.
In the following Sections we then focus on the analysis of the \lit
production by different stellar sources. In particular in \S~4 we
consider the role of intermediate-mass AGB stars (hereafter IMS-AGB),
presenting new $^7$Li yields obtained from the AGB models by
Frost~(1997).  We reanalyze the production of \lit by novae (\S~5),
SNII (\S~6), and low-mass RGB stars (\S~7), on the basis of recent
model calculations and observations.  Finally, in \S~8 we discuss and
summarize our results.

\section{Observational data}

\subsection{\lit abundances in field dwarfs}

The main source of our compilation is the recent study by
Fulbright~(2000) who  determined \lit, Fe, and other element abundances
for 168 halo and disk stars. Both stellar parameters and abundances of
\lit and Fe were inferred by carrying out a self-consistent LTE
analysis, providing a homogeneous sample of both Pop.~II and Pop.~I
stars.  The survey of Fulbright does not extend to very low or very
high metallicities; therefore we complemented it for [Fe/H] $<-2$ with
the compilation of Bonifacio \& Molaro~(1997). At [Fe/H] $> -0.4$ we
included a sample of warm ($T_{\rm eff} \geq 5700$~K), unevolved F-type
stars taken from Balachandran~(1990) and Lambert, Health, \&
Edvardsson~(1991). We note that, whereas cooler stars have certainly
undergone Li depletion (unless very young), considering only stars
warmer than 5700~K allows having a subsample of minimally depleted
stars which trace the upper envelope of the \lit vs. [Fe/H]
distribution.  Surveys of \lit among Pop.~II stars have also been made
by e.g.  Rebolo et al.~(1988), Thorburn~(1994), Ryan, Norris, \&
Beers~(1996), Spite et al.~(1996), Ryan et al.~(2001). Different
datasets and/or analysis methods (in particular the use of different
effective temperature calibrations) lead in general to a different
abundance scale: therefore, including \lit abundances from many
different sources in the same compilation may introduce a spurious
(i.e. not real) dispersion in the \lit vs [Fe/H] distribution.  For
this reason, we decided to base our compilation of \lit abundances for
Pop.~II stars on the minimum number of studies allowing the full
coverage of the entire metallicity range. We mention that
Fulbright~(2000) derived an average offset between his and Bonifacio \&
Molaro's (1997) effective temperatures $\Delta T_{\rm eff}=-38 \pm
20$~K, which should not introduce major systematic differences in the
inferred \lit abundances. There are several stars in common between the
two studies: for all of them the agreement in \lit abundances is indeed
very good (within $\sim 0.1$~dex).

We did not include in our list stars evolved off the main sequence
(hereafter MS) and/or stars cooler than 5700~K; both groups of stars
most likely did not preserve their original \lit content due to either
post-MS dilution or to MS \lit depletion and thus are not adequate to
trace the upper envelope of the \lit vs.  metallicity distribution.  We
also did not include \lit abundances for T Tauri stars, since they may
be affected by several problems such as spectral veiling, NLTE, or,
more generally, by the effects of a ``disturbed stellar photosphere"
(see e.g. Duncan~1991).  Instead, we included in our compilation the
average abundance of warm (i.e. F--type) stars in young open clusters
with $-0.1 <$[Fe/H]$<0.1$.  These stars have undergone minimal
pre-MS or MS \lit depletion and thus their abundance can be considered
as representative of the present ISM \lit content. Several surveys of
\lit among open cluster stars (see e.g. Jeffries~2000 for a recent
review) have shown that Li-undepleted stars in all the investigated
clusters share the same abundance $\log \epsilon(^7{\rm Li})=3.2\pm
0.1$ (e.g., Randich et al.~1997)\footnote{Although \lit abundances as
high as $\log \epsilon(^7{\rm Li})=4$ have been derived for some
T~Tauri stars, Magazz\`u, Rebolo, \& Pavlenko~(1992) and Mart\'\i n et
al.~(1994) concluded that abundances of pre-MS stars (including both
classical and weak-lines T Tauri stars) as a whole indicate an initial
\lit abundance for Pop.~I stars $\log\epsilon\rm(^7{\rm Li}) =3.1$.}.

In the upper panel of Figure~1 we plot the \lit abundance vs.
metallicity for the stars of our sample.  A few features in this figure
warrant detailed comments.  As far as the \lit plateau is concerned,
recent studies have inferred slightly different values for the plateau
value, due -- as mentioned above -- to different types of analysis and,
in particular, to possible offsets in the zero-point of the effective
temperature scale (see e.g., Spite et al.~1996). In our model of
Galactic evolution we have generally adopted as initial \lit abundance
the value determined by Bonifacio \& Molaro (1997) for the Spite
plateau, i.e.  $\log \epsilon(^7{\rm Li})_0=2.24\pm 0.012$, under the
implicit assumption that the Spite plateau represents the primordial
(Big-Bang) abundance.  We mention in passing that the mean abundance of
the stars in the compilation of Fulbright~(2000) with [Fe/H]$\leq -1$
is $\log \epsilon(^7 {\rm Li})_0 =2.24\pm 0.095$ in excellent agreement
with the estimate of Bonifacio \& Molaro~(1997). 
We also ran our model starting with an initial \lit abundance
higher by a factor 2 than the Spite plateau value (see below), following
the idea that Pop.~II stars may have depleted a certain amount of Li in
the course of their evolution.

Figure~1 also suggests that the rise of \lit abundance from the Spite
plateau may be smoother than claimed in previous studies. In
particular, a few stars are present in the survey of Fulbright~(2000)
which have intermediate metallicities ($-0.7 \leq$ [Fe/H] $\leq -0.45$)
and \lit abundances 0.2--0.3 dex (i.e. $>1\sigma$) above the plateau
but not as high as $\log\epsilon (^7{\rm Li})\simeq 3$. However, the caveat 
is that only a few stars define the upper envelope at intermediate
metallicity and we cannot exclude that these stars have suffered some
Li depletion. 

A sharp increase to abundances as high as the present ISM
abundance may occur at metallicities between [Fe/H]$=-0.45$ and
$-0.35$, if the abundances for the few stars which trace the upper
envelope of the distribution at those metallicities are correct.  These
stars are from the survey of Balachandran (1990), but we mention that Li abundances 
derived by Balachandran are systematically higher than those of Boesgaard \& Tripicco
(1986), with differences in the range 0.2 -- 0.6 dex.
For the case of HR 8315 (HD 206901) L\`ebre et al.~(1999) inferred a metallicity
[Fe/H]$=-0.3$ and a \lit abundance $\log \epsilon(\rm Li)=2.9$ to be compared with
[Fe/H]$=-0.37$ and $\log \epsilon(\rm Li)=3.05$ determined by Balachandran~(1990).  
We also notice that photometric iron abundances for some of the
intermediate metallicity  and Li-rich stars observed by Balachandran (1990) are at 
least 0.2 dex higher than the spectroscopic
values inferred by Balachandran (see the last two columns of Table~1 in
Balachandran~1990), suggesting again that \lit abundance $\sim 3$ is
reached only at higher metallicities (i.e. [Fe/H]$\geq -0.3$). 
Determinations of $^7$Li abundances in this critical metallicity range are
certainly needed in order to better constrain the \lit vs. [Fe/H] upper
envelope.

\subsection{\lit abundances in clusters}

In the lower panel of Figure~1 we show the abundance of \lit vs.
Galactic age using \lit data for stars in Galactic open and globular
clusters. We also show the meteoritic \lit abundance
($\log\epsilon(^7{\rm Li})=3.31\pm 0.04$, Anders \& Grevesse~1989), the
interstellar \lit abundances determined in the line-of-sight toward
$\rho$ Oph (\lit/H $\gtrsim 2.4\times 10^{-10}$, Lemoine et al.~1993),
and toward $o$ and $\zeta$ Per (\lit/H $=(9.8\pm 3.5)\times 10^{-10}$
and $(12.2\pm 2.2)\times 10^{-10}$, Knauth et al.~2000).  For young
open clusters (ages $< 1$~Gyr) the average abundance measured in warm
($T_{\rm eff} \geq 6000$~K), undepleted cluster MS stars, is plotted in
the Figure. The data were taken from the following sources: Soderblom
et al.~(1999) for NGC~2264; Randich et al.~(1997) for IC~2602; Randich
et al.~(2001) for IC~2391; Mart\'\i n \& Montes~(1997) for IC~4665;
Randich et al.~(1998) for Alpha Persei; Soderblom et al.~(1993) for the
Pleiades; Thorburn et al.~(1993) for the Hyades. On the other hand, the
initial abundance for older clusters was estimated from stars at (or
just evolved off) the turn-off in the 2~Gyr clusters NGC~752 (Hobbs \&
Pilachowski~1986), NGC~3680 (Randich, Pasquini, \& Pallavicini~2000)
and from the abundances of the two components of the tidally locked
binary S1045 in the solar-aged M~67 cluster (Deliyannis et al. 1994;
Pasquini, Randich, \& Pallavicini~1997).  Finally, globular clusters
(GCs) are represented by a box that shows the ranges in \lit and age
estimates.  More specifically, GCs absolute ages are the subject of a
long standing debate; in the lower panel of Figure~1 we used the recent
determination by Carretta et al.~(2000), who inferred an average age of
$12.9 \pm 2.9$~Gyr based on {\it Hipparcos} parallaxes for local
subdwarfs. We mention, however, that ages as large as 16--17 Gyr cannot
be ruled out.

Abundances of \lit were derived for unevolved stars at the turn-off in
three clusters only, namely, NGC~6397 (Pasquini \& Molaro~1996), 47 Tuc
(Pasquini \& Molaro~1997), and M~92 (Boesgaard et al.~1998); whereas
the average abundances for the three clusters are close to the value of
the Spite plateau, stars in the same cluster show a significant
dispersion in \lit (see discussion in Boesgaard et al.~1998) and, in
particular, a star is found in M~92 with a \lit content significantly
higher than the plateau and the other cluster stars.  Consequently, we
think that the use of a box is more appropriate to indicate the
position of GCs in Figure~1.

\section{The Galactic chemical evolution model}

The model of Galactic chemical evolution adopted in this work is
described in detail by Ferrini \& Galli (1988), Galli \& Ferrini (1989)
and Ferrini et al. (1992). The same model was adopted by Galli et
al.~(1995), Travaglio et al.~(1999) and Travaglio et al.~(2001), to
study the evolution of the light elements D and $^3$He, and the
evolution of the heavy elements from Ba to Pb, respectively. We briefly
recall here the basic features of the model.

The Galaxy is divided into three zones, halo, thick disk, and thin
disk, whose composition of stars, gas (atomic and molecular) and
stellar remnants is computed as functions of time up to the present
epoch $t_{\rm Gal}=13$~Gyr. The thin disk is formed from material
infalling from the thick disk and the halo. The formation of the Sun in
the thin disk takes place 4.5~Gyr ago, i.e. at epoch
$t_\odot=8.5$~Gyr.  The star formation rate in the three zones is not
assumed {\em a priori}, but is obtained as the result of
self-regulating processes occurring in the molecular gas phase, either
spontaneously or stimulated by the presence of other stars.

In models of Galactic chemical evolution it is customary to separate
the contribution of stars in different mass ranges to the enrichment of
the ISM. Our model follows the evolution of ({\em i}\/) single low- and
intermediate-mass stars ($0.8~M_\odot\leq M \leq M_\star$) ending their
life as He or C-O white dwarf, ({\em ii}\/) binary systems able to
produce type {\small I} supernovae, and ({\em iii}\/) single massive
stars ($M_\star\leq M\leq 100~M_\odot$), the progenitors of type {\small
II} supernovae. The value of $M_\star$ depends on metallicity; we
assume $M_\star=6$~\ms\/ for [Fe/H]$\le -0.8$, and $M_\star=8$~\ms\/
otherwise (see for references Tornamb\`e \& Chieffi~1986).  The adopted
initial mass function (hereafter IMF) is discussed in Ferrini et
al.~(1992).  Stellar nucleosynthesis is treated according to the matrix
formalism introduced by Talbot \& Arnett (1973).  Specific yields are
taken from calculations by Woosley \& Weaver (1995) and Thielemann,
Nomoto, \& Hashimoto (1996), for type II and type I supernovae,
respectively.

\section{The production of \lit by hot bottom burning process}

Hot bottom burning (hereafter HBB) occurs in intermediate mass AGB
stars when the bottom of the convective envelope of the star reaches
the top layers of the hydrogen burning shell. Sackmann \&
Boothroyd~(1992) showed that the Cameron-Fowler mechanism
(Cameron~1955; Cameron \& Fowler~1971) is very effective at producing
\lit, and can quantitatively account for the \lit abundance measured by
Smith \& Lambert (1990) in the super-Li rich AGB stars in the
Magellanic Clouds. We present here a summary of results for \lit
production in AGB stars of various masses and metallicities, based on
the evolutionary models of Frost~(1997) and synthetic models
using the technique of Forestini \& Charbonnel~(1997). We then 
discuss the sensitivity of \lit
enrichment to the mass-loss prescriptions during the latest phases of
AGB evolution.

\subsection{Li production from 4, 5, 6 $M_\odot$ AGB models at different
metallicities}

Frost~(1997) computed the evolution of 4, 5, and 6 \msb models each
with metal content $Z = 0.02$, 0.008, and 0.004, appropriate to the
solar neighborhood, and to the Large and Small Magellanic Clouds,
spanning the evolution from the MS through the end of the AGB phase.
Each case has been the subject of a detailed nucleosynthesis study
using a modified post-processing code (Cannon~1993) including 74
species and over 500 reactions. Some of the results of these models
have been already published (e.g. Frost et al.~1998, Lattanzio \&
Forestini~1999); in this paper we present in detail an analysis of
the production of $^7$Li.

Our present approach is to use the stellar models of Frost~(1997) as the
basis of synthetic models computed with the code of Forestini \&
Charbonnel~(1997). In this way we force the synthetic code to follow
the results of the stellar models (accuracy is within 10\%), yet we are
free to vary those parameters which have no significant feedback on the
evolution.  In particular, for the aim of this paper, we investigated
the mass-loss rate at the end of the AGB phase (see next Section for
detailed discussion).  We do not expect modest
changes in the mass-loss rate during the latest AGB phase to have
a significant effect on the structural evolution of the star, since
the variations in the mass-loss rate only affect the very late phases
of stellar evolution. Nevertheless, we caution against placing too much
belief in the quantitative results for the extreme cases discussed below.

The surface \lit abundances during the AGB lifetime in the stellar
envelope for the nine models considered here is presented in Figure~2.
Note that the abundance of \lit at the beginning of the AGB phase
varies from one model to another as a result of previous evolution.
Fig.~2 is a $3\times 3$ grid figure with mass increasing along the
$x$-axis and $Z$ decreasing along the $y$-axis: in this way we expect
HBB to increase for increasing $x$-axis (mass) or increasing $y$-axis
(decreasing $Z$). The most extreme HBB is for the top right plot, i.e.
$M = 6$~\msb model with $Z = 0.004$, where the maximum \lit production
lasts for the longest time ($\sim 10^5$~yr). The maximum \lit
abundances we obtain range around \nli $\simeq 4$, independent of
stellar mass and metallicity, and only depending on the occurrence of
HBB.  The models with $M = 4$~\msb and $Z = 0.02$, 0.008 show no
HBB and consequently no \lit production. In fact, \lit destruction
in these two models proves that their convective bottom temperatures
are high enough to burn \lit through $^7$Li$(p,\alpha)^4$He channel, but
not enough to produce $^7$Be (that decays to $^7$Li, Cameron-Fowler mechanism) 
by $^3$He burning. Conversely the 6 \msb
models, with their substantial envelopes, begin HBB immediately on the
AGB and hence they produce a large amount of \lit, but the high
envelope temperature also assists in the destruction of \lit. 
Finally, Figure~2
shows that in all models, as the evolution of the star proceeds, the
surface \lit abundance decreases again more or less rapidly,
depending on the mass and metallicity. This indicates that the initial
reserves of $^3$He have been consumed in the production of \lit, and
this \lit is now itself being destroyed (without replenishment) by
proton captures.

There are observational indications that some Galactic C-stars (stars
with C/O $> 1$) are also super-Li rich stars (\nli $\ge 3$, see e.g.
Abia et al.~1991, Abia, Pavlenko, \& de Laverny~1999, Abia \&
Isern~2000).  Therefore, we also examined the \lit vs. C/O trends for
the nine cases presented above.  The models of 
Frost~(1997) use the algorithm for the third dredge-up (i.e. the
penetration of the convective envelope into the partially He-burnt zone
after each thermal pulse), as described by Frost \& Lattanzio~(1996),
and include the entropy adjustment by Wood~(1981). After each third
dredge-up episode the $^{12}$C abundance is increased.  Hence third
dredge-up increases the ratio C/O (by adding C to the stellar
envelope), and HBB (where it is active) decreases it again as it
transforms C into N.  Figure~3 shows the C/O ratio vs. \lit for the
nine models under consideration.  In some cases HBB prevents the
formation of a C star, while in other cases it merely delays the
formation to a time when HBB has stopped and third dredge-up continues
(see Frost et al.~1998 for more details). The condition C/O $> 1$
together with high \lit values are obtained only for the case $Z =$
0.004 and $M =$ 4 \msb and for a very short phase for the case with the
same metallicities and $M =$ 5 \ms.  In particular, the 4 \msb model is
a Li-rich star (with \nli $\sim$ 4) and C-rich star for a period of
about 10$^5$ yr.  Moreover, all the other cases show high \lit
abundances together with O-rich envelopes, in agreement with
observations of AGB stars in the Magellanic Clouds (see e.g. Smith et
al.~1995) and in the Galaxy (see e.g. the recent work from Arellano
Ferro, Giridhar, \& Mathias~2001).

The range of $M_{\rm bol}$ covered by the nine models presented here
($-7\le M_{\rm bol} \le -6$) during the Li-rich phase is in agreement
with bolometric magnitudes of the super-Li rich stars observed in the
Magellanic Clouds (Smith et al.~1995). Concerning Galactic super-Li
rich C stars, only few estimates of their $M_{\rm bol}$ are available. 
Abia et al.~(1991), based on the distance determinations by Claussen et
al.~(1987), estimated $M_{\rm bol} \simeq -5$ for WZCas and WXCyg (with
\nli $=5.0$ and 4.7, respectively). More recently, Abia \&
Isern~(2000), based on {\it Hipparcos} parallaxes, inferred $M_{\rm
bol}=-$6.44 and \nli$=$4.8 for WZCas, and $M_{\rm bol}=-$4.35 and
\nli$=4.4$ for WXCyg. However, a caveat is that the latter is based on a
parallax $\pi=-1.41 \pm 1.98$, which is unreliable.
Two additional Galactic
C-rich super-Li rich stars have been presented by Abia et al.~(1991):
IYHya ($M_{\rm bol}=-6.2$ and \nli$=5.4$), and TSgr ($M_{\rm bol}=-5.8$
and \nli$=4.2$). Given the uncertainties, and excluding the not
reliable measurement of $M_{\rm bol}$ for WZCas by Abia \&
Isern~(2000), we conclude that the $M_{\rm bol}$ determinations for
the Galactic super-Li rich stars are still consistent with the $M_{\rm
bol}$ reached by our AGB models during the super-Li rich phase.  We
also notice that the highest abundance reached by our models for \nli
is $\sim 4$, while higher values are inferred from observations.

\subsection{Superwind in AGB stars: consequences for Li chemical enrichment}

There are observational indications that toward the end of their
evolution, AGB stars (both of low- and intermediate-mass) can
experience a short phase of extremely rapid mass-loss, called {\it
superwind} (van der Veen, Habing, \& Geballe~1989; Schr\"oder, Winters,
\& Sedlmayr~1999). In this phase periods of heavy mass-loss may be
interspersed with much longer periods of lower mass-loss rates,
possibly powered by He shell flashes. These stars are often surrounded
by a dense circumstellar envelope and some of them are no longer
visible at optical wavelengths. Evidence for a superwind has been found
in CO observations of C stars (e.g. Olofsson~1993), of K giants (de La
Reza et al.~1997; Castilho et al.~1998), and in post-AGB objects that
show concentric rings around the central object (see e.g. Crabtree \&
Rogers~1993; Latter et al.~1993; Riera et al.~1995; Klochkova et
al.~1999; Bl\"ocker et al.~2001).

From the theoretical point of view, the term of superwind has been
coined (Renzini~1981) to describe the heavy, final tip-AGB mass-loss
($\ge 10^{-5}$~$M_\odot$~yr$^{-1}$), which is required to form a
planetary nebula of typically tenths of a solar mass within several
10$^4$ yr.  The idea of a star terminating its AGB life with a
superwind has been also elaborated by Bowen \& Willson~(1991).  These
authors showed that all stars undergo the superwind phase as a result
of an increase in scale height and density at the condensation
radius as the star evolves toward the tip of the AGB.  Analytical
expressions to describe this phase of high mass-loss have been
presented by different authors, e.g.  by Vassiliadis \& Wood~(1993,
hereafter VW93), Bl\"ocker~(1995), Salasnich, Bressan, \&
Chiosi~(1999). VW93 computed the effects of thermal pulses on mass-loss
rates and suggested that a star may undergo several superwind phases
and that more massive stars will stay longer in the superwind phase
than lower mass stars. For the AGB models discussed in this Section the
superwind prescription follows the formula of VW93 (without the
correction for masses above 2.5 \ms)
\begin{equation}
\log\left(\frac{dM}{dt}\right)   = -11.4 + 0.0125 P,
\end{equation}
where the mass-loss rate is in $M_\odot$~yr$^{-1}$ and $P$ 
is the pulsation period in days. Note that the mass-loss
rate is truncated at
\begin{equation}
\frac{dM}{dt}=\frac{L}{cv_{\rm exp}},
\end{equation}
where $L$ is the stellar luminosity,
corresponding to a radiation-pressure driven wind.  Again from VW93
we take the wind expansion velocity (in km~s$^{-1}$)
\begin{equation}
v_{\rm exp} = -13.5 + 0.056 P,
\end{equation}
ranging between 3.0 and 15.0 km~s$^{-1}$.

The \lit yields are very sensitive not only on the extent of the HBB,
but also on the mass-loss rate prescriptions. As first
qualitatively pointed by Abia, Isen, \& Canal (1993), it is not clear if
during the super-Li rich phase AGB stars are able to inject $^7$Li into 
the interstellar medium by the high mass-loss rate before $^7$Li is
depleted in the atmosphere. We analyze this point with our models,
and we plot in Figure~4 the surface \lit abundance as a function of the
mass-loss rate for the nine models
described in the previous Section. The figure shows that only in a few
cases does the star achieve very high mass-loss rates
($>10^{-5}$~\ms~yr$^{-1}$) when the surface \lit abundances is still
high (\nli $> 3$).  When this happens the overall \lit yield is
positive, or only slightly negative.  In the first three columns of
Table~1 we report the \lit yields (in \ms) obtained with these models,
showing that only in the case of 4 \msb and $Z = 0.004$ the \lit yield
is positive.

\begin{table}
\begin{center}
TABLE 1\\
{\sc \lit yields (in solar masses) from intermediate-mass AGB stars}
\begin{tabular}{c|ccc|ccc}   
\hline
\hline
     &   &  ({\it a}\/) &  &   &  ({\it b}\/) &  \\
\hline
     & $Z=0.004$ & $Z=0.008$ & $Z=0.02$ & $Z=0.004$ & $Z=0.008$ & $Z=0.02$ \\
\hline
4~\msb & $ 1.1\times 10^{-8}$ & $-1.7\times 10^{-8}$ & $-4.3\times 10^{-8}$ & $1.5\times 10^{-8}$ & $-1.3\times 10^{-8}$ & $-3.3\times 10^{-8}$ \\
5~\msb & $-1.1\times 10^{-8}$ & $-1.9\times 10^{-8}$ & $-4.0\times 10^{-9}$ & $4.5\times 10^{-8}$ & $ 7.3\times 10^{-8}$ & $ 2.2\times 10^{-8}$ \\
6~\msb & $-1.3\times 10^{-8}$ & $-2.5\times 10^{-8}$ & $-1.3\times 10^{-8}$ & $4.1\times 10^{-8}$ & $ 7.2\times 10^{-8}$ & $ 7.7\times 10^{-8}$ \\
\hline
\hline
\end{tabular}
\end{center}
\noindent ({\it a}\/) Mass-loss rate from VW93 \\
\noindent ({\it b}\/) Mass-loss rate from VW93 increased by a factor 50 \\
\end{table}

Another way to investigate the sensitivity of \lit yields to the choice
of the mass-loss rate is shown in Figure~5, where we plot the surface
\lit abundance as a function of the total stellar mass (so that the
models evolve from right to left).  Note that for the models with $Z =$
0.02 we assumed an initial (i.e. at the time when the star formed)
$\log \epsilon (^7{\rm Li}) \simeq 3.3$, compatible with the meteoritic
abundance; this value scales with $Z$ for the other metallicities. In
order to get a positive yield of \lit the star's mass should decrease
substantially while $\log \epsilon \rm (^7{\rm Li}) > 3$--4.  Hence,
although the 5 and 6 \msb models experience  HBB, the 5 \msb loses
about 1~\msb when $\log\epsilon(^7{\rm Li}) \simeq 4$, whereas the 
6 \msb loses most of its mass when $\log \epsilon (^7{\rm Li})$ is 
3.5 or lower. Hence the yield from 5 \msb model,
although still negative, is smaller in its absolute value than the
6 \msb model. That is, the 5 \msb model does not destroy as much \lit 
as the 6 \msb model, due to the large amount it returns to the ISM,
although the overall yield is still negative, with respect to the
initial abundance.  At a lower value of $Z$ ($\sim 0.008$) where
typical interior temperatures are higher, both the 5 \msb and 6 \msb
models clearly show HBB, and there is a small amount of HBB at 4~\ms,
although $\log \epsilon (^7{\rm Li})$ never exceeds 2. Again, most of
the mass-loss for these models occurs when the surface $\epsilon(^7{\rm
Li})$ is less than the initial value, and hence the overall yields are
negative (see Table~1).  Finally, at $Z = 0.004$ the higher temperature
means that there is significant \lit production even at 4~\ms. This is
particularly strong in the more massive models, but this does not
correspond to a positive \lit yield because the \lit is destroyed
before the high mass-loss begins. In contrast, the 4~\msb model has its
HBB delayed, and toward the end of its evolution the increased
mass-loss begins when the surface \lit is higher, and hence the overall
yield is higher. Also note that the high values found for $\log \epsilon
\rm (^7{\rm Li})$ in the 5 and 6~\msb models do not last until the
mass-loss reaches appreciable values. By the time that the mass-loss
begins in earnest, the surface \lit abundance is below the initial
value. For the 4~\msb case, however, there is about a half solar mass
of material lost when $\log \epsilon (^7{\rm Li})$ is above 3. A
conclusion from these results is that the higher \lit yields come  from
the lower mass stars: in fact, from those stars which just start HBB
when the high mass-loss rate begins.

Due to the fact that \lit yields are very sensitive to the mass-loss
choices, we have run each of the models presented above using different
mass-loss rates.  Since several observations of OH/IR stars (i.e.
O-rich AGB stars that exhibit OH masers, Wilson \& Barrett~1972) with
infrared excesses (i.e. high mass-loss rates, see e.g. Bl\"ocker et al.~2001) 
and $P \sim 400$--500 days are now available (see e.g. Lewis~2000 for a 
recent survey of
OH/IR IRAS sources), we forced the mass-loss to start at shorter
periods.  In order to obtain $dM/dt \ge 10^{-5}$~$M_\odot$~yr$^{-1}$
with $P \simeq 500$ days the VW93 prescription has to be increased of a
factor of $\sim 50$.  In Table~1 (columns 5, 6, 7) we list the yields
obtained with this modification to the VW93 prescription. The surface
\lit abundance for the different models vs. stellar mass is also shown
in Figure~5 for comparison. Both the table and the figure indicate
that, when using this mass-loss rate prescription, the \lit yields for
most of the cases are positive.

The chemical evolution model described in \S~3 has been run using
the Li yields given in Table~1 (first three columns): we obtain that
intermediate-mass AGB stars contribute to the solar system \lit
abundance for $\sim 14$\%.  Under these conditions, IMS-AGB stars can
contribute only  a small fraction to the \lit solar composition and
they do not seem able to reproduce the rapid increase of \lit in the
Galactic disk.  When we introduce in
the GCE model the \lit yields obtained the modified VW93 mass-loss
prescription (Table~1, last three columns) and we find that the
contribution of intermediate-mass AGB stars to the solar \lit
composition increases by a few percent ($\sim 6$\%). In Figure~6
({\it upper panel}) we compare the observed \nli~vs [Fe/H] distribution
with the predictions of the model including the modified VW93 mass-loss
prescription.

We also notice that we limited the IMS-AGB upper mass to 6~\msb. This is 
due to the fact that we did not yet run models for 7 and 8 \msb AGB stars.. 
Nevertheless we extrapolate the 6 \msb yields to 7 and 8
\msb models to test the 4--8 \msb IMS-AGB mass range. We found that
the results change only by few percent, due to the lower weight on the IMF
of the 7 and 8 \msb with respect to the 4 \msb stars.

As a test on the results presented in this Section, we also run several
AGB models with a Reimers mass-loss (Reimers~1975) and different values
of the Reimers parameter $\eta$ ($\eta=1$, 5, 10 for all masses and
metallicities). We find that, even for the higher $\eta$ values, the super-Li
rich phase is reached with a mass-loss rate smaller than $\sim 5\times
10^{-6}$~$M_\odot$~yr$^{-1}$. Therefore, the resulting \lit yields are 
very close to those obtained with the VW93 standard case (shown in the 
first three columns of Table~1). However, we stress that large variations 
in the parameter $\eta$ may induce significant feedback on the AGB evolution,
and the results of our synthetic models may not be completely
reliable.

\subsection{Galactic Li enrichment: HBB in 3 \msb AGB stars with low
$Z$?}

It is commonly believed that HBB only occurs for masses greater than
about 4 \ms.  This is not strictly true: the temperature at the base of
the envelope has also a strong dependence on the metallicity of the
star.  When the temperature at the base of the deep convective envelope
is larger than $\sim 2\times 10^7$~K, $^7$Be is efficiently produced
and the surface \lit abundance increases. For the models presented
here, HBB occurs with initial masses $\ge 5$~$M_\odot$ at each
metallicities and $\ge 4$  $M_\odot$ for $Z = 0.004$. In principle, at
lower metallicities the inner envelope becomes hot enough to start HBB
at lower masses, but models of lower metallicities through the AGB
phase are rare.

Preliminary calculations for low values of $Z$ have been presented by
Lattanzio et al.~(2001).  They found that, for $Z=0.0001$, envelope
temperatures as high as $\sim 2\times 10^7$~K are reached for masses as
low as 2.5~\ms. Because of the shape of the IMF we would thus expect
these stars to contribute substantially to \lit production, via HBB.
In addition, as shown above, the maximum abundances obtained in the
models discussed in this paper are $\log\epsilon (^7{\rm Li}) \simeq 4$,
independently on the stellar mass and metallicity. Therefore, under
these preliminary indications, we tested the sensitivity of $^7$Li
enrichment when we extend the HBB mass range to 3--6~\msb at low
metallicities ($Z < 0.004$). For $M < 4$~$M_\odot$ we adopted the same
yields of the 4~$M_\odot$ model presented above (detailed models are in
preparation), and for $M \ge$ 4$~$\msb we used the yields obtained
with VW93 increased by a factor of 50 (see Table~1 and discussion 
in the previous Section. The result for GCE of \lit is shown in the lower panel of
Figure~6. There is a significant increase of the \lit enrichment mostly
due to the higher weight of these stars on the IMF, as well as an
important increase of the \lit contribution of these stars at 
the epoch of the solar system formation  ($\sim 40$\%).

\section{The production of \lit by novae}

Novae can in principle contribute to the \lit enrichment of the ISM via
the production and subsequent decay of $^7$Be (i.e. Cameron-Fowler
Beryllium mechanism). This process can be investigated in detail only
with the help of hydrodynamical simulations of the accretion and
explosive phases of evolution of a nova system. Early works by Arnould
\& N\o rgaard~(1975),  Starrfield et al.~(1978), Shara~(1980), and Iben
\& Tutukov~(1984) showed that \lit can be produced in considerable
amounts during nova outbursts. Subsequent studies by Boffin et
al.~(1993) and Coc et al.~(1995) pointed out the sensitivity of the
results to the adopted nuclear reaction network and the treatment of
convection between the accreted envelope and the underlying white dwarf
core.  A systematic analysis of the chemical composition of the ejecta
of both CO and ONe novae has been recently presented by Hernanz et
al.~(1996) and Jos\'e \& Hernanz~(1998, hereafter JH98) by means of
hydrodynamical simulations following both the accretion and the
explosion phase. Their results can be summarized as follows:  ({\em
i}\/) the production of $^7$Be is weakly dependent (logarithmically) on
the initial $^3$He concentration in the envelope, for fractional
abundances of $^3$He larger than the solar value; ({\em ii}\/) the
final \lit abundance depends rather sensitively on the chemical
composition of the envelope, which, in turn, depends on the composition
of the underlying core:  typically, the \lit abundance by mass is in
the range $\sim 10^{-6}$--$10^{-5}$ in the case of a CO white dwarf,
and $\sim 10^{-7}$--$10^{-6}$ in the case of ONe white dwarfs; ({\em
iii}\/) the predicted ejected mass during a nova outburst is $\sim
10^{-5}$~\ms.

It should be stressed that the amount of mass ejected during a nova
outburst predicted by hydrodynamical models ($\sim 10^{-5}$~\ms) is
systematically lower than the value observationally determined in a
small sample of nova systems, peaked around $\sim 10^{-4}$~\ms\ (see
e.g. Della Valle~2000 for a summary of results). However, the envelope
masses inferred from observations are highly uncertain.  In addition,
the usual assumption that the ejected shells are almost homogeneously
filled in (filling factor $\sim 0.1$--1) has been challenged by recent
observations of the Nova T Pyx (Shara et al.~1997), suggesting values
of the filling factor in the range $10^{-2}$--$10^{-5}$.  Incomplete
knowledge of the mass of nova ejecta is the most serious limitation to
a quantitative evaluation of the role of novae as \lit sources in the
Galaxy.

Crude estimates of the total amount of \lit synthesized by Galactic
novae have been given by Starrfield et al.~(1978), Hernanz et
al.~(1996), and JH98. They all consistently show that novae can account
for $\sim 10$\% of the Galactic \lit content. The contribution of novae
to the Galactic evolution of \lit has been considered by D'Antona \&
Matteucci~(1991), Matteucci et al.~(1995), Romano et al.~(1999), in the
framework of models of Galactic chemical evolution.  However, a
detailed study of the role of novae as \lit producers necessarily
suffers from several uncertainties.  On one hand, the relevant
quantity, i.e. the amount of \lit ejected into the ISM during a nova
outburst, is the product of two poorly constrained factors: the \lit
abundance and the total mass ejected.  On the other hand, the
evaluation from first principles of the nova rate in the Galaxy
requires the accurate knowledge of a number of physical processes and
quantities which are neither theoretically nor observationally well
determined, such as  the white dwarf cooling timescale, the fraction of
binary stars, and the fraction of binary stars that end up as a nova
system.

In this Section, we estimate the {\it upper limit} to the amount of
\lit synthesized by novae in the Galaxy, in particular deriving
approximate analytical expressions for the abundance of \lit as a
function of the gas metallicity.  In the framework of the ``closed
box'' model for the solar neighborhood (see e.g.  Tinsley~1980), it is
easy to predict the evolution of the \lit mass fraction $X_7(\mu)$  as
a function of the gas fraction $\mu$ in the Galaxy.  Let's define a {\em
nova mass ejection rate} $\varphi_{\rm N}(t)=M_{\rm ej}\nu_{\rm N}(t)$
(in \ms~yr$^{-1}$), where $\nu_{\rm N}(t)$ is the nova rate (in
yr$^{-1}$), and assume that the rate of mass ejection by novae is
proportional to the star formation rate, $\varphi_{\rm
N}(t)=\alpha\psi(t)$, with $\alpha$ independent of time.  Indicating
with $\langle X_{\rm 7}\rangle_{\rm ej}$ the average mass fraction of
\lit in nova ejecta, and $X_{\rm 7in}$ the initial (cosmological) \lit
abundance, we obtain 
\begin{equation} 
\label{x7mu} 
X_7(\mu)=X_{\rm 7in}\mu^{R/(1-R)}+\frac{\alpha \langle X_{\rm 7}\rangle_{\rm
ej}}{R}[1-\mu^{R/(1-R)}], 
\end{equation} 
where $R\simeq 0.21$ according
to Galli et al.~(1995) is the {\em stellar returned fraction} over the
Galactic lifetime.  The first term in eq.~(\ref{x7mu}) represents the \lit
destruction by astration, the second term the \lit production by
novae.  It is convenient to eliminate the gas fraction $\mu$ in favour
of the metallicity of the gas $Z$, given by 
\begin{equation}
\label{zmu} 
Z(\mu)=-\frac{P_Z}{1-R}\ln\mu, 
\end{equation} 
where $P_Z\simeq 7.9\times 10^{-3}$ (Galli et al.~1995) is the {\em
metal production factor}.

Notice that the contribution of novae to the evolution of \lit in the Galaxy
depends almost linearly, at late times, on the combination $\alpha   
\langle X_{\rm 7}\rangle_{\rm N}$. The value of $\alpha$, being constant,
can be estimated at the present time $t=t_{\rm Gal}$,
\begin{equation}
\label{al}
\alpha=\frac{M_{\rm ej}\nu_{\rm N}(t_{\rm Gal})}{\psi(t_{\rm Gal})}
\simeq 2\times 10^{-4}
\left(\frac{M_{\rm ej}}{3\times 10^{-5}~\mbox{\ms}}\right)
\left(\frac{\nu_{\rm N}(t_{\rm Gal})}{35~\mbox{yr}^{-1}}\right)
\left(\frac{\psi(t_{\rm Gal})}{5~\mbox{\ms~yr}^{-1}}\right)^{-1},
\end{equation}
where we have adopted the value of the nova rate recently proposed by
Shafter~(1997) and the star formation rate given by Metzger~(1988).  As
for $M_{\rm ej}$ and $\langle X_{\rm 7}\rangle_{\rm ej}$ we have
adopted the average values given by Romano et al.~(1999) for
$t>4.5\times 10^7$~yr, based on the results of JH98: $M_{\rm ej}\simeq
3\times 10^{-5}$~$M_\odot$ and $\langle X_{\rm 7}\rangle_{\rm ej}
\simeq 3\times 10^{-6}$. Clearly for $\alpha\sim 10^{-4}$ the role of
novae in the Galactic evolution of \lit is marginal, the \lit abundance
predicted by the simple model at $t=t_\odot$ ($X_7\simeq 1\times
10^{-9}$) being only $\sim 10$\% of the meteoritic value
($X_{7\odot}\simeq 9.3\times 10^{-9}$). This result is in agreement
with the order-of-magnitude estimates by Starrfield et al.~(1978),
Hernanz et al.~(1996), and JH98.

Several approximations have been made in the derivation of the results
of this Section. However, they are likely to lead to an overestimate of
the predicted \lit abundance. Consider for instance the assumption of
Instantaneous Recycling Approximation (IRA) for novae.  Owing to the
long evolutionary timescales necessary to produce a nova system, the
ratio $\alpha$ was considerably smaller in the past (see D'Antona \&
Matteucci~1991, Romano et al.~1999), and the approximation of a nova
rate proportional to the SFR, with the proportionality constant
$\alpha$ estimated at the present time, results in an overestimate of
the past nova rate and therefore the \lit abundance.  Neglecting infall
also results in an overestimate of the \lit abundance predicted by the
model, if the infalling material has primordial composition and
therefore acts as a diluting effect on the disk \lit abundance.  The
same is true neglecting the delay of about $\sim 1$~Gyr
estimated by D'Antona \& Matteucci~(1991) between the onset of star
formation in the Galaxy and the birth of the first nova system.

\begin{table}
\begin{center}
TABLE 2\\
{\sc Production of $^{13}$C, $^{15}$N, $^{17}$O, and $^{7}$Li by novae}
\begin{tabular}{ccclcl}
\hline
\hline
& $X_\odot$ & $\langle X\rangle_{\rm ej}$~({\it a}\/) &  & $\langle X\rangle_{\rm ej}$~({\it b}\/) &  \\
\hline
$^{13}$C & $3.6\times 10^{-5}$ & $1.0\times 10^{-1}$ & 100\% (assumed) & $2.4\times
10^{-2}$ & 100\% (assumed) \\
$^{15}$N & $3.6\times 10^{-6}$ & $1.5\times 10^{-2}$ & 150\% & $5.3\times
10^{-2}$ & 220\% \\
$^{17}$O & $3.4\times 10^{-6}$ & $1.1\times 10^{-2}$ & 120\% & $3.0\times
10^{-2}$ & 130\% \\
\hline
{\bf $^{7}$Li} & $9.3\times 10^{-9}$ & $3.0\times 10^{-6}$ & {\bf 12\%} &
$9.2\times 10^{-7}$ & {\bf 14\%} \\
\hline\hline
\end{tabular}
\end{center}
\noindent ({\it a}\/) CO novae (JH98 models)\\
\noindent ({\it b}\/) ONe novae (JH98 models)\\
\end{table}

Finally, another independent argument against a {\it dominant}
contribution of novae to the production of $^7$Li is the constraint on the
abundance of isotopes like $^{13}$C, $^{15}$N and $^{17}$O produced
copiously by novae according to JH98. The constraint that we discuss below 
is independent of any specific model
of chemical evolution.  In Table~2 we show the average mass fraction of
these isotopes in nova ejecta, assuming the yields computed by JH98 for
CO and ONe novae (third and fifth column), together with their
abundances in the protosolar material (first column), representative of
the ISM composition at $t=t_\odot$ (from Anders \& Grevesse~1989, and
Grevesse, Noels \& Sauval~1996). In the fourth and sixth
column we show the contribution of novae to the solar abundance of each
element {\it assuming that novae are the only producers of} $^{13}$C.
This is obviously not the case, since it is well known that low- and
intermediate-mass stars produce significant amounts of this isotope
(see e.g. Palla et al.~2000 and references therein). We see from
Table~2 that even under this favourable assumption on the role of novae
in the chemical enrichment of the Galaxy, their contribution to the
solar Li abundance cannot be greater than $\sim$ 10\%.

\section{The production of Li by SNII}

Woosley et al.~(1990) and Woosley \& Weaver~(1995, hereafter WW95)
advanced the idea that production of \lit via the so called {\it
neutrino process} in SNII could account entirely for the Solar System
\lit abundance. Timmes, Woosley, \& Weaver~(1995), using a full grid of
SNII models of various masses and metallicities, predicted a lower \lit
production rate by the $\nu$-process than Woosley et al.~(1990),
concluding that SNII contribute about one-half the solar \lit abundance
(see also Matteucci et al.~1995 on this point).

With our model of GCE, assuming the \lit yields
of WW95, we obtain the results shown in Figure~7. Starting with an
initial \lit composition of $X_{7{\rm in}} = 1.0\times 10^{-10}$, we
found that SNII can account for $\sim 40$\% of the meteoritic \lit
content, in agreement with Timmes et al.~(1995), and Matteucci et
al.~(1995). In order to show the exact value of the metallicity at
which SNII in our GCE model start to contribute significantly, we also
show in Fig.~7 the contribution of SNII computed with initial $X_{7{\rm
in}} =0$.

We should notice, however, that the SNII Li yields computed by WW95
have been questioned in recent theoretical studies of hydrodynamics and
rotation in SNII (see e.g. Langer et al.~1999, and Heger, Langer, \&
Woosley~2000).  In particular, Langer et al.~(1999) showed that the
inclusion of rotational mixing of the envelope of massive MS stars
(later supposed to end as SNII) drastically reduces the amount of
$^3$He present in the stellar interior. In non-rotating massive stars
this isotope is found to be neither produced nor destroyed. As shown by
Langer et al.~(1999) for a 15 \msb rotating model, $^3$He production
factor is 0.1\% of the correspondent case in a non-rotating WW95
model.  Since \lit is largely produced by the
$^3$He($\alpha$,$\gamma$)$^7$Li reaction, the inclusion of rotation in
massive star models is expected to drastically reduce the production of
\lit, with respect to the non-rotating cases. Thus, the results shown
in Fig.~7 should be considered as upper limits for the contribution of
SNII to the \lit enrichment in the Galaxy.

\section{The production of \lit by deep mixing in low-mass giants stars}

\begin{table}
\begin{center}
TABLE 3\\
{\sc \lit yields (in solar masses) by deep mixing in low-mass giant stars}
\begin{tabular}{cccc}
\hline
\hline
$M$ & mixing speed & $Z=Z_\odot$ & $Z=0.001$\\
(\ms) & (\msb yr$^{-1}$) & & \\
\hline
1.0 $^{({\it a}\/)}$ & $10^{-3}$ & $3.8\times 10^{-10}$ & $5.1\times
10^{-10}$ \\
    & $10^{-4}$ & $2.0\times 10^{-9}$  & $7.6\times 10^{-10}$ \\
    & $10^{-5}$ & $1.0\times 10^{-8}$  & $3.6\times 10^{-10}$ \\
\hline
2.5 $^{({\it b}\/)}$ & $10^{-3}$ & $2.1\times 10^{-10}$ & $3.0\times
10^{-10}$ \\
    & $10^{-4}$ & $7.6\times 10^{-10}$ & $6.7\times 10^{-10}$ \\ 
    & $10^{-5}$ & $6.0\times 10^{-9}$  & $2.1\times 10^{-9}$ \\
\hline
\hline
\end{tabular}
\end{center}
\noindent ({\it a}\/) SB99 models.\\
\noindent ({\it b}\/) Our extrapolations from ({\it a}\/).\\
\end{table}

Pop.~I low mass giant stars ($M <$ 4 $M_\odot$) are thought to produce Li
both in the RGB and AGB phases. More specifically, low mass stars, below 
2.5 $M_\odot$,
produce \lit in the RGB and in the AGB by a deep mixing process such
as explored by  Wasserburg, Boothroyd, \& Sackmann~1995, and Sackmann
\& Boothroyd~1999, hereafter SB99.  This process is sometimes called
cool bottom processing. We prefer to refer to this as ``deep mixing''
because it is more indicative of the physics involved: the cool bottom
of the envelope does not play any role. The nuclear
processing occurs at the top of the H-shell and this is facilitated by
mixing from the bottom of the convective envelope down to the deeper
layers of the H-shell.  This mechanism needs to be further explored to
see if it can be efficient also in the mass range 2.5 $< M/M_\odot <$
4.0.

According to stellar evolution models, RGB stars should be
characterized by a relatively low \lit content: stars that leave the MS
undepleted in \lit, after the first-dredge up dilution has occurred,
are expected to have \lit abundances of the order of \nli~$\sim
1.5-1.9$ (Iben 1965, 1967a,b).  Since most stars destroy \lit on the MS,
their \lit content when they reach the RGB and in immediately
subsequent phases should be much lower than the above values.  Most
field and cluster RGB stars show in fact very low \lit abundances
($-1.5<$ \nli $<0$), even lower than predicted by the models. An
extra--mixing mechanism has been proposed to explain those low
abundances (see e.g. Charbonnel, Brown, \& Wallerstein~1998; Charbonnel
\& Balachandran~2000, and references therein). However, after the first
discovery by Wallerstein \& Sneden (1982) of a Li-rich RGB (i.e. a RGB
star with a \lit abundance higher than model predictions), several
other Li-rich giants have been found (for recent surveys, see Castilho
et al.~1998; Jasniewicz et al.~1999). Note that, although the number of
presently known Li-rich RGB stars is relatively high, they represent
only 1--2~\% of all the giants with \lit measurements (Wallerstein \&
Sneden~1982; Gratton \& D'Antona~1989; Pilachowski, Hudek, \&
Sneden~1990; Pallavicini et al.~1990; Fekel \& Balachandran~1993).
Some of these Li-rich RGB stars have abundances even higher than the
present ISM value (de la Reza \& da Silva~1995; Balachandran~2000).

Various suggestions have been made to explain the Li-rich giants with
\nli $\ge 2$ (for stars with $1 <$ \nli $< 2$ a fresh \lit production
is not necessary since their \lit abundance is consistent with the
first dredge-up values). The high \lit content of these giants may be
related to external processes (Alexander~1967; Gratton \&
D'Antona~1989; Siess \& Livio~1999) or to internal processes 
such as the production of fresh \lit (Fekel
\& Balachandran~1993; de la Reza, Drake, \& da Silva~1996; SB99).
As shown by SB99, 
\lit can be created in low-mass RGB stars via the
Cameron-Fowler mechanism associated with a deep-mixing below the
convective envelope. This internal circulation (possibly driven by 
stellar rotation) transports envelope
material into the outer wing of the H-burning shell, where it undergoes
partial nuclear processing, and then back to the envelope. SB99 choosed 
the free parameters of
their model to match the low value of $^{12}$C/$^{13}$C observed in RGB
stars. In order to produce the required additional $^{13}$C, the advected material 
must reach temperatures high enough that $^3$He is burned,
resulting in the creation of $^7$Be via
$^3$He($\alpha$,$\gamma$)$^7$Be. If extra-mixing is slow, $^7$Be is
destroyed via $^7$Be($p$,$\gamma$)$^8$B($e^+$,$\nu$)$^8$Be or
$^7$B($e^-$,$\nu$)$^7$Li, and any \lit produced from $^7$Be electron
captures is immediately burned up via $^7$Li($p$,$\alpha$)$^4$He.
However, for higher mixing speed $^7$Be can be transported out to
cooler regions before electron capture takes place and the stellar
envelope becomes enriched in \lit.  The production of \lit in RGB
stars is dependent on the mixing-speed, i.e.  the stream mass flow
rate. SB99 discussed the range of values of this mixing-speed and
argued that it must be slower than the velocity of convection in RGB or
AGB ($\sim$ 1 \msb yr$^{-1}$), while the streams must move faster than
the speed with which the H-shell burns its way outward. We show in
Table~3 how the different values for the mixing-speed can influence the
\lit production (see also SB99 for more details on the 1 \msb model). 
It is important to notice that, as discussed by e.g.
Charbonnel~(1994) and Boothroyd \& Sackmann~(1999), the mass range in
which the deep mixing is active is $\sim$ 1.0--2.5~\ms. This is due to
the fact that for low mass stars ($\le 2.5$~\ms), the H-burning shell
catches up to and erases the discontinuity while the star is still on
the RGB; for higher masses ($>$ 2.5~\ms) the star leaves the RGB
before this can take place.

We use the models by SB99 to estimate the contribution of low-mass RGB
stars to the chemical evolution of the Galaxy.  The \lit production
phase on the RGB is short ($\sim 10^5$~yr) compared to the total red
giant lifetime ($\sim 5\times 10^7$~yr), and since the typical RGB
mass-loss rate is rather low, they are not expected to contribute
significantly to the \lit enrichment in the ISM. On the other hand,
during the TP-AGB phase, because of the higher mass-loss rate, the
contribution to the \lit ISM enrichment can be substantial.  For these
reasons, and since SB99 do not present specific \lit predictions for
the AGB phase, in this work we make the following assumptions. We
consider the SB99 \lit predictions at the end of the RGB phase (when
$\log(L/L_\odot)\simeq 3.4$) for 1~\msb at two different metallicities
($Z=Z_\odot$ and $Z=0.001$) and for three different cases of
mixing-speed ($10^{-5}$, $10^{-4}$, $10^{-3}$~\ms~yr$^{-1}$). To obtain
the yield of \lit from these models we multiply the \lit abundances
by the total mass ejected, e.g. for 1 \msb model we used $M_{\rm ej}
\simeq 0.45$~\msb (see e.g.  Weidemann~1984 for the initial-final mass
relationship).  We assume a mass range of 1--2.5 \ms, in agreement with
the above discussion, and since the authors showed only the results for
the 1 \msb model, we interpolate the \lit production by deep mixing
in the mass range 1~\msb -- 2.5~\ms.  To derive the 2.5~\msb \lit
yields we follow SB99 assuming that all $^3$He is converted in \lit,
and that the abundance of $^3$He in stars scales as $M^{-2}$
(Schatzman~1987). The \lit yields obtained for 1~\msb and 2.5~\ms,
with different metallicities and different mixing speed, are shown in
Table~3.  In our calculations, in order to derive an upper limit to the
\lit production, we adopt the highest \lit abundance predicted by
the SB99 models, i.e. that obtained with the model with 10$^{-5}$ \msb
yr$^{-1}$ mixing speed.

The contribution of low-mass giant stars to the Galactic evolution of
\lit is shown in Figure~8. These stars enrich \lit up to $X_7 \sim
2.2\times 10^{-9}$ at the epoch of formation of the Sun, i.e.  $\sim
24$\% of the solar value. Again, another more efficient source is
required to explain the ISM \lit abundance. As in the case of SNII
(Fig.~7) we show for comparison the contribution from these stars to
\lit enrichment starting with a zero initial \lit composition.

\section{Conclusions}

We have addressed the long-standing problem of the Galactic chemical
evolution of \lit with the aim of clarifying the role of the different
stellar contributions. We have analyzed four possible stellar sources
of \lit:  SNII, novae, low-mass giants and IMS-AGB stars.  For low-mass
giants, novae, and SNII we have critically examined the available \lit
yields and discussed the possible extrapolations when no model
predictions were available. In the case of SNII and novae we
re-examined the \lit yields in the light of recent hydrodynamical
computations.  For IMS-AGB stars we presented and used here new results
for nucleosynthesis calculations based on  evolutionary
AGB models by Frost~(1997). In particular, we discussed how the
interplay between the HBB process and a phase of high mass-loss before
the evolution off the AGB may constitute a key process for \lit
enrichment in the Galaxy. Although we are not yet able to quantify the
contribution of IMS-AGB stars, we have explored different realistic
possibilities in terms of mass-loss rate and mass range for the HBB
process. 

The work presented here is summarized in Figure~9, where we compare the
results of our model of chemical evolution with the sample of
observational data presented in Fig.~1. The predicted \lit abundance
resulting from all stellar sources considered in this paper is plotted
in Fig.~9 vs. metallicity (upper panel) and vs. time (lower panel).
For IMS-AGB we used a VW93 mass-loss increased by a factor of 50 
(as discussed in \S~4.2), and we have also taken into account the results 
obtained for the 3~\msb AGB model presented in \S~4.3.

As we discussed in Sect.~2.1, we also ran our model with an initial \lit abundance
higher by a factor 2 than the Spite plateau value. The result is 
shown for comparison in Fig.~9. Notice that the predicted \lit abundance
at the time of formation of the Sun is virtually the same in both cases.
In the lower panel of Fig.~9
we plot the individual contributions from the different stellar
sources, with the exception of the contribution from SNII since 
we believe
that the inclusion of rotation and mixing in the supernova models leads
to a drastic reduction of the \lit yields (see \S~6).  
Note that, from an observational point of view, it is critical
to determine the metallicity where the rise of the \lit abundance from
the Spite plateau occurs; the available data indicate that the rise
from the plateau occurs at metallicities [Fe/H] between $\simeq -1$ and
$-0.8$, but additional data in the metallicity range $-1 <$ [Fe/H] $<
-0.3$ are needed in order to better constrain this value.

Our main conclusions are the following:

({\it i}\/) In the light of the available stellar models, small
contributions to the meteoritic \lit abundance come from novae ($\sim
10$\% or less, see \S~5) and low-mass giant stars ($\sim 20$\%,
see \S~7). As for SNII, we predict a contribution to the solar
system \lit abundance of $\sim 40$\% with the standard yields by
Woosley \& Weaver~(1995), and a contribution less than 10\% if the
results of the latest hydrodynamical simulations of the supernova
explosion are taken into account (see \S~6).  Spallation reactions
between ISM and cosmic-ray nuclei can provide an additional at least
$\sim 10$--20\% of the \lit abundance in the solar system (see e.g.
Lemoine et al.~1998).

({\it ii}\/) Figure~9 shows that even with our best-fit model we are
not able to reproduce the meteoritic abundance, although both cluster
and ISM abundances are fitted fairly well.  Our best-fit model is in
good agreement with the observed \nli~vs. [Fe/H] distribution up to
[Fe/H] $\simeq -0.4$. This indicates that the metallicity at which the
rise from the Spite plateau occurs is consistent with the [Fe/H] values
at which the IMS-AGB stars contribute to the chemical enrichment of the
ISM. We remind that the observational data for [Fe/H] $>-0.4$ are mostly
taken from Balachandran~(1990) (see discussion in \S~2); nevertheless, 
as the comparison between photometric and spectroscopic metallicities 
shows, part of these stars might have higher [Fe/H] values.

({\it iii}\/) Our best-fit model and, more specifically, the
contribution from IMS-AGB, is based on two major assumptions:  first
the high mass-loss phase at the end of the evolution of these stars
must start {\it earlier} on the AGB than the standard predictions by
VW93. This is supported by recent IR observations,
discussed in \S~4.  Second, we also need a contribution to \lit
through HBB production from stars with 3 \msb $\le M \le 4$~\msb and
low metallicities (see preliminary model results for $M<4$~\msb in
Lattanzio et al.~2001).  Should these two assumptions be not valid, the
contribution from intermediate-mass AGB would be much lower. We stress
however that, since neither novae, nor SNII, nor low-mass giant stars
seem to produce enough \lit to account for the present-day abundance,
we believe that intermediate-mass AGB stars remain at present the best
candidates as \lit factories.

Our results and conclusions are strongly based on the AGB models
discussed in \S~4 (see also Frost~1997). A different view has been 
recently expressed by Ventura et al.~(2000). They presented a grid 
of IMS-AGB models of different
masses and tested the sensitivity of \lit yields to different mass-loss
rate prescriptions on the basis of a comparison with AGB stars in the
Magellanic Clouds.  Ventura et al.~(2000) concluded that, with their
mass-loss calibration, IMS-AGB stars do not contribute significantly to
the \lit enrichment of the ISM. As they properly noticed, details of
\lit production depend on the input parameters of the stellar model,
mainly the treatment of convection. In addition,
Ventura et al.~(2000) focused their work on the analysis of the
strength of the mass-loss rate during the AGB phase. We instead
followed a different approach, testing different times during the AGB
phase at which the high mass-loss rate starts, and looking for the
consequences on \lit yields. As we demonstrated in \S~4,
anticipating by few thermal pulses the beginning of the superwind phase
can have significant impact in the mass of \lit ejected.

Work in progress includes an extension of the grid of AGB models to
lower masses ($M < 4$~\ms) and lower metallicities ($Z < 0.004$) (see
preliminary results in Lattanzio et al.~2001), in order to  analyze the
possibility of \lit production via HBB in these stars.  Another
interesting point that we just mentioned in this paper, is a
re-analysis of the contribution of GCR to the production of \lit (Valle et
al., in preparation). Finally, new \lit data for a
statistically significant sample of stars in the critical metallicity
range $-1<$ [Fe/H] $< -0.3$ are being analyzed; the data will provide
stringent observational constraints on the Galactic \lit abundance in a
metallicity range where IMS-AGB stars are expected, on the basis of
this work, to give their main contribution.

\acknowledgements

We thank S. Shore for interesting discussions and comments on the manuscript. C.T. also
thanks R. Gallino for useful suggestions and encouragements during the
development of this work.  We also thanks the anonymous referee for very
usefull comments that improved our paper.
The research of C.T., S.R. and D.G. is partially supported by grants
COFIN98 and COFIN2000.

\clearpage

\begin{figure}
\plotone{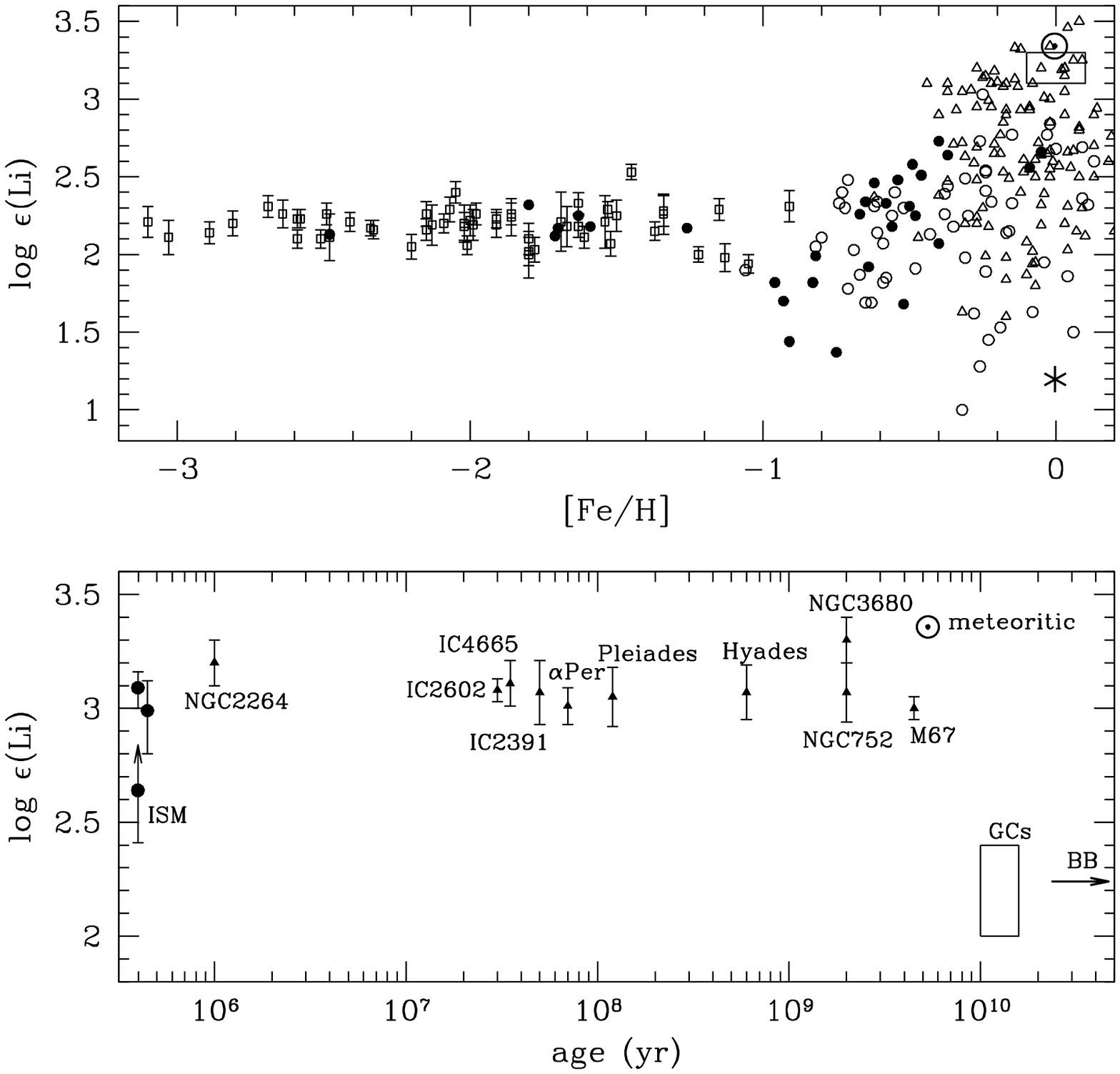}
\figcaption{{\it Upper panel}: $\log\epsilon (^7{\rm Li})$ vs. [Fe/H]
for field stars. Observations are from Fullbright~(2000) ({\it filled
circles}), Bonifacio \& Molaro~(1997) ({\it open squares}),
Balachandran~(1990) ({\it open triangles}), Lambert et al.~(1991) ({\it
open circles}). The box represents the region occupied by stars in open
clusters with undepleted \lit. The meteoritic and the (solar) photospheric
values are also shown by a {\it dotted circle} and an {\it asterisk},
respectively.  {\it Lower panel}: $\log\epsilon (^7{\rm Li})$ vs. age
for Galactic open clusters ({\it filled triangles}), and globular
clusters ({\it box}). We also include different measurements of the ISM
value ({\it filled circles}), the meteoritic value ({\it dotted
circle}), and the cosmological \lit abundance adopted in this work (see
text for references).}
\end{figure}
\clearpage

\begin{figure}
\plotone{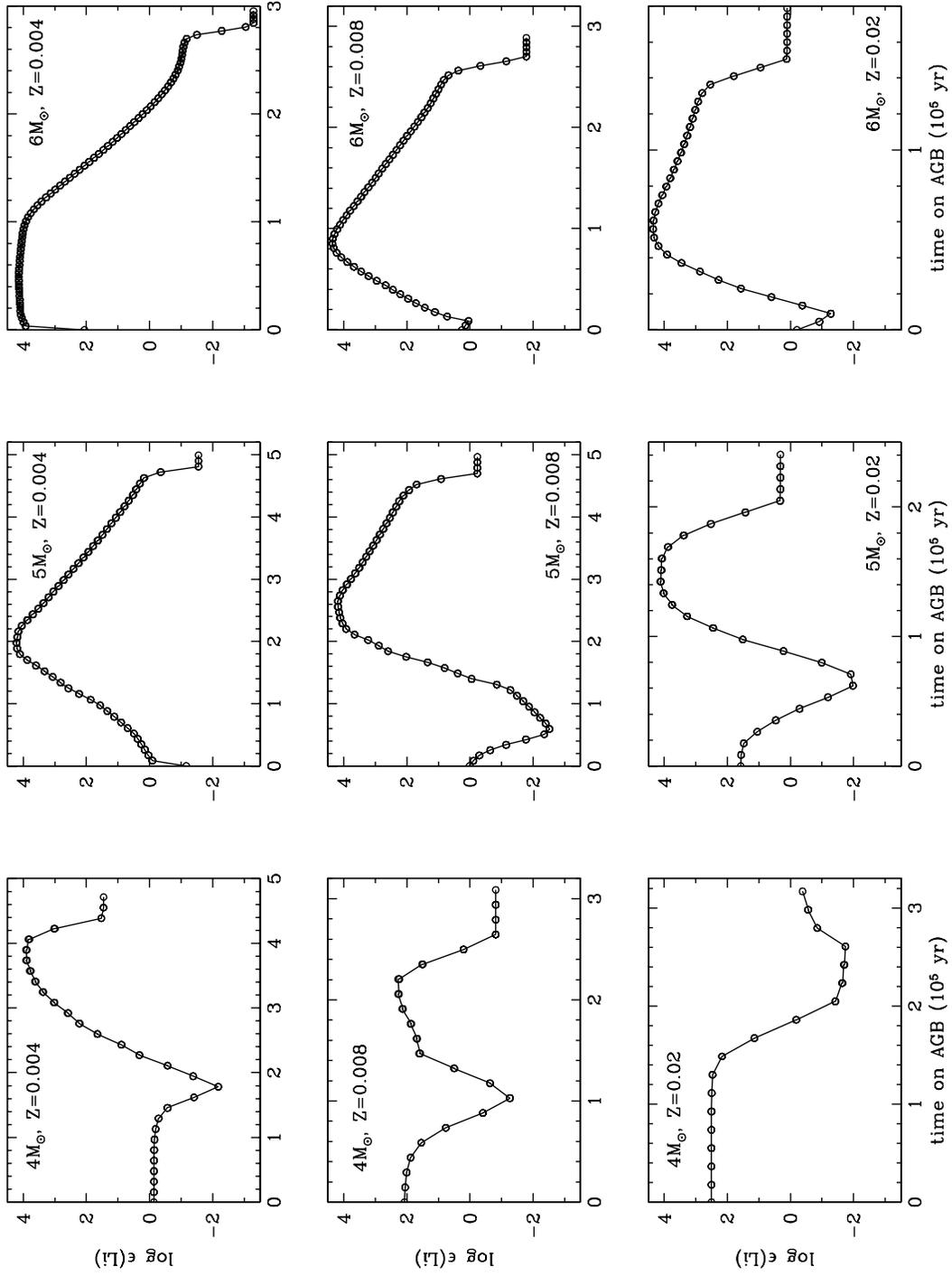}
\figcaption{Surface \lit abundance in the envelope of 4, 5, 6 $M_\odot$ 
stars with $Z = 0.004$, 0.008, 0.02, vs. time on the AGB. Each
circle represents a thermal pulse in the model.} 
\end{figure}
\clearpage

\begin{figure}
\plotone{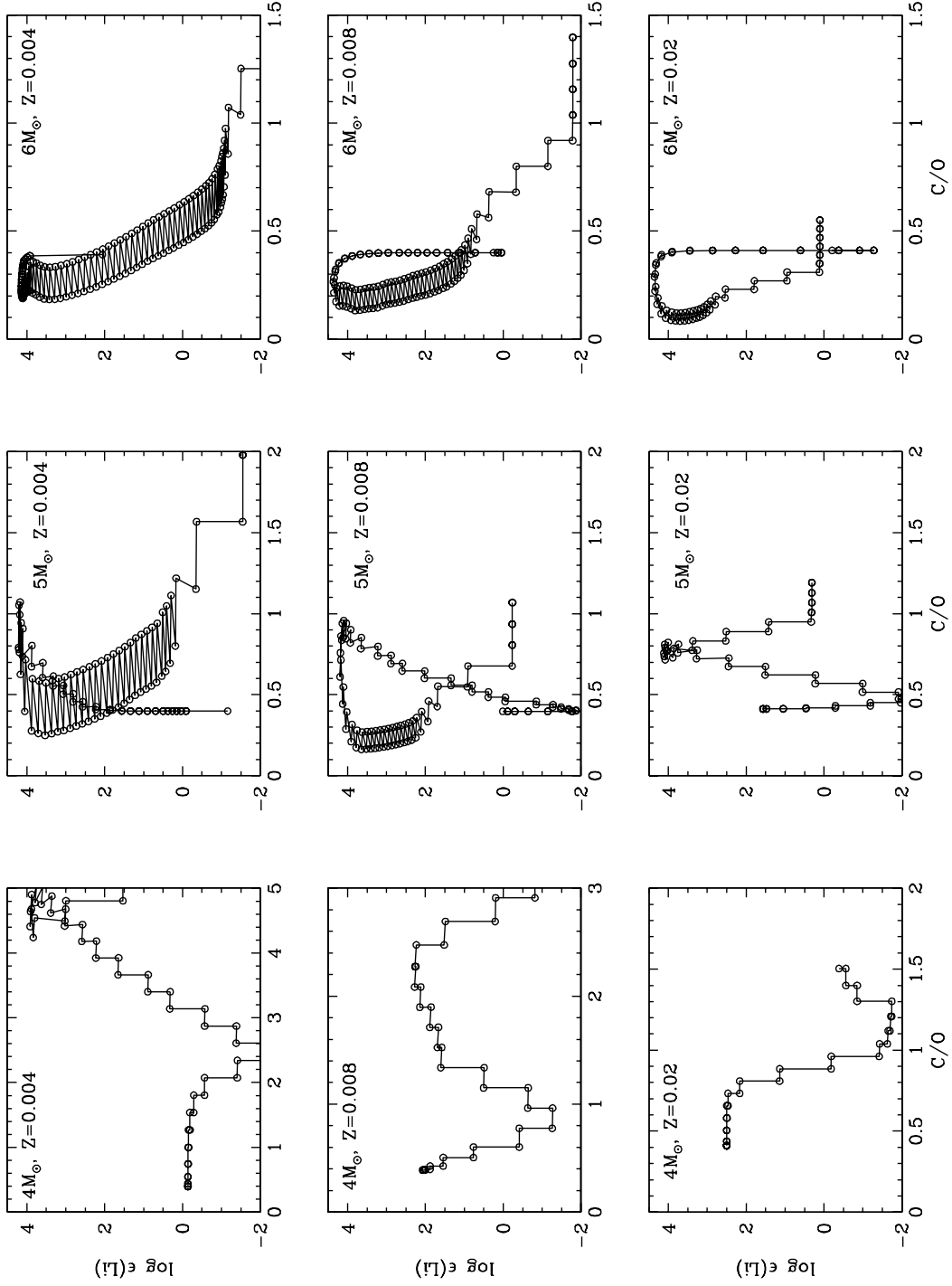}
\figcaption{Surface C/O ratios vs. \lit surface abundances for the nine
AGB models described in the text. Symbols are as Fig.~2.}  
\end{figure}
\clearpage

\begin{figure}
\plotone{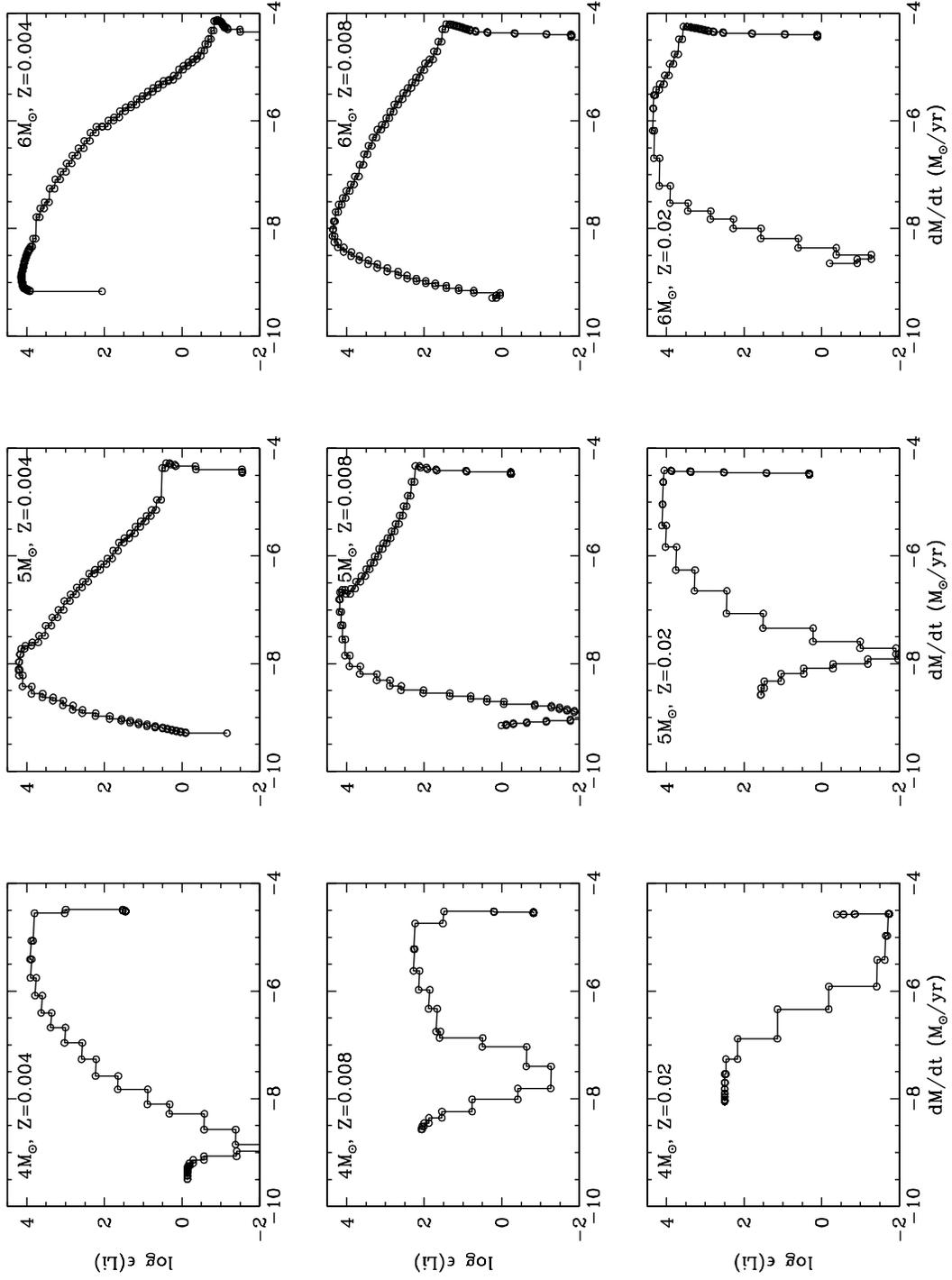}
\figcaption{Surface \lit abundance vs mass-loss rate (in \ms~yr$^{-1}$) 
for the nine AGB models described in the text, following the mass-loss
prescriptions of Vassiliadis \& Wood~(1993).  Symbols are as Fig.~2.}
\end{figure}
\clearpage

\begin{figure}
\plotone{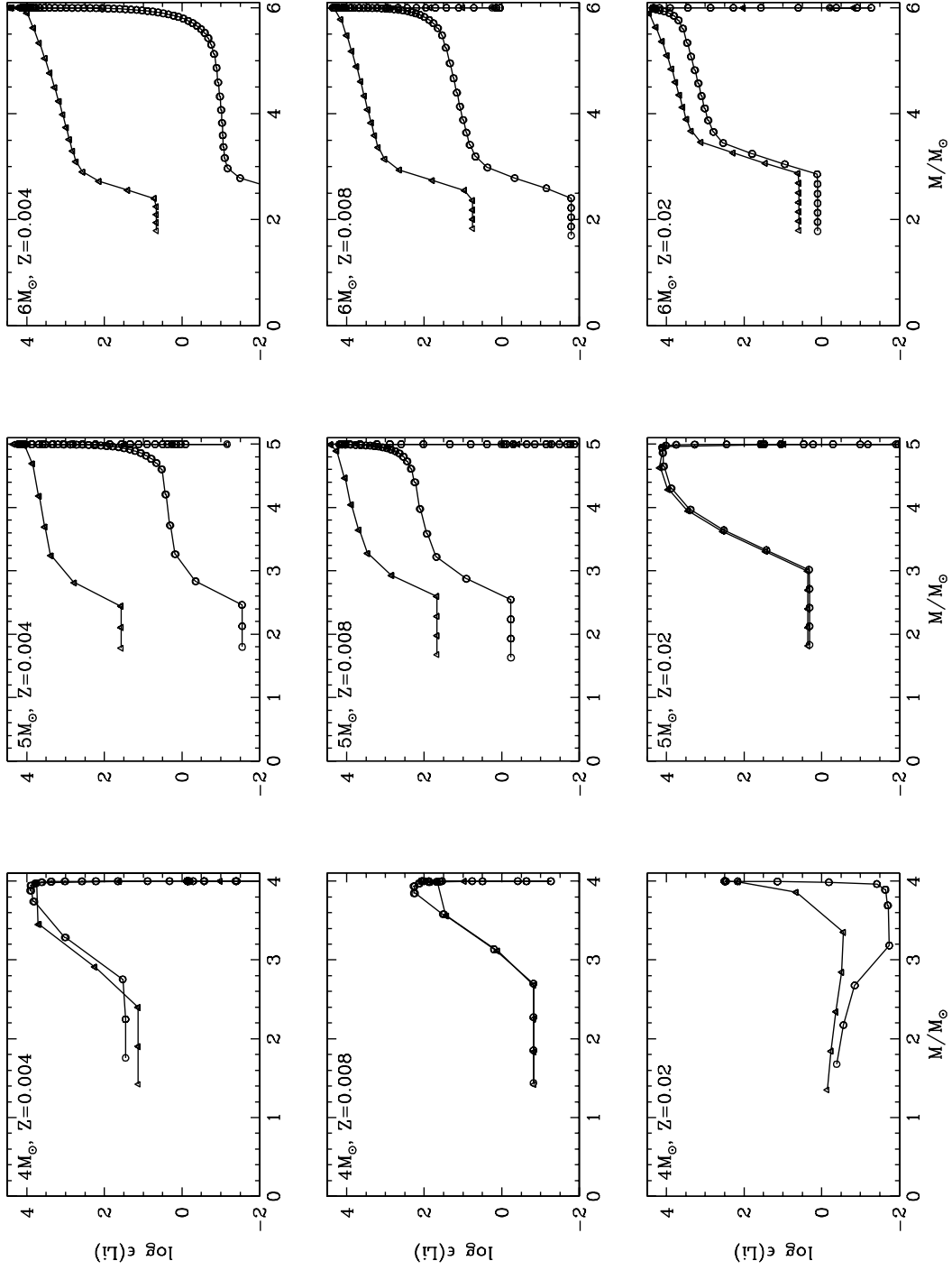}
\figcaption{Surface \lit abundance vs. stellar mass for the AGB models
discussed in the text. The {\it open circles} are for the Vassiliadis
\& Wood~(1993) mass-loss, the {\it open triangles} are for the
Vassiliadis \& Wood~(1993) mass-loss increased by a factor 50. Each
open circle and open triangle represents a thermal pulse.}
\end{figure}
\clearpage

\begin{figure}
\plotone{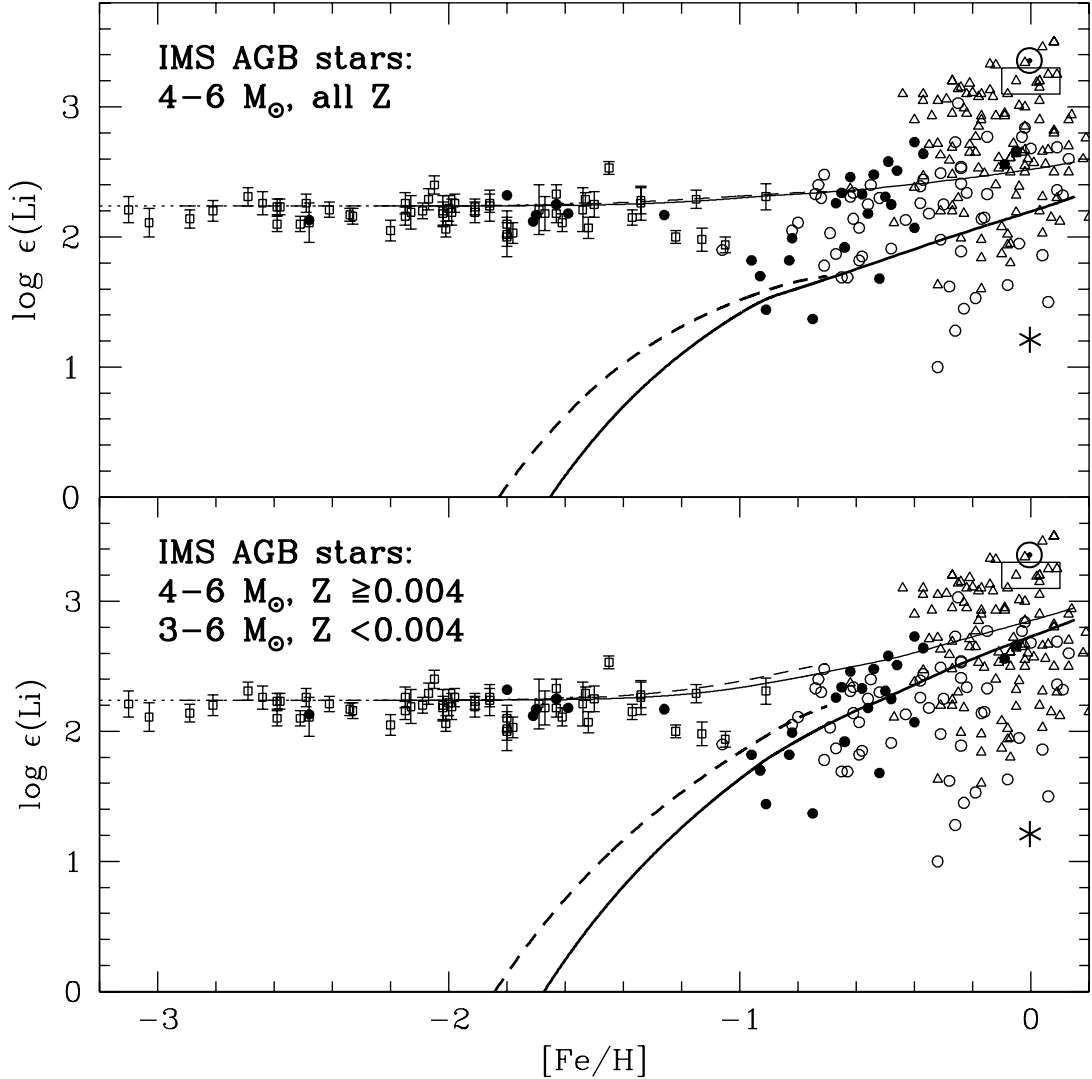}
\figcaption{Contribution of IMS-AGB stars to the Galactic 
evolution of \lit, for two different choices of the IMS-AGB mass range:
4--6 \msb ({\it upper panel}), and 4--6 \msb for $Z \ge 0.004$ and
3--6 \msb with $Z < 0.004$ ({\it lower panel}) (see also text for
details). In both cases we adopt the mass-loss rate prescription of
Vassiliadis \& Wood~(1993) with modification (see text).
Symbols are as in Fig.1 (upper panel).  Lines refer to the GCE model
results for halo ({\it dotted}), thick disk ({\it dashed}), and thin
disk ({\it solid}). Thick lines show the \lit evolution obtained
with a zero initial abundance of \lit, to emphasize the contribution of
IMS-AGB stars in the thick disk ({\it dashed line}) and in the thin
disk ({\it solid line}).}
\end{figure}
\clearpage

\begin{figure}
\plotone{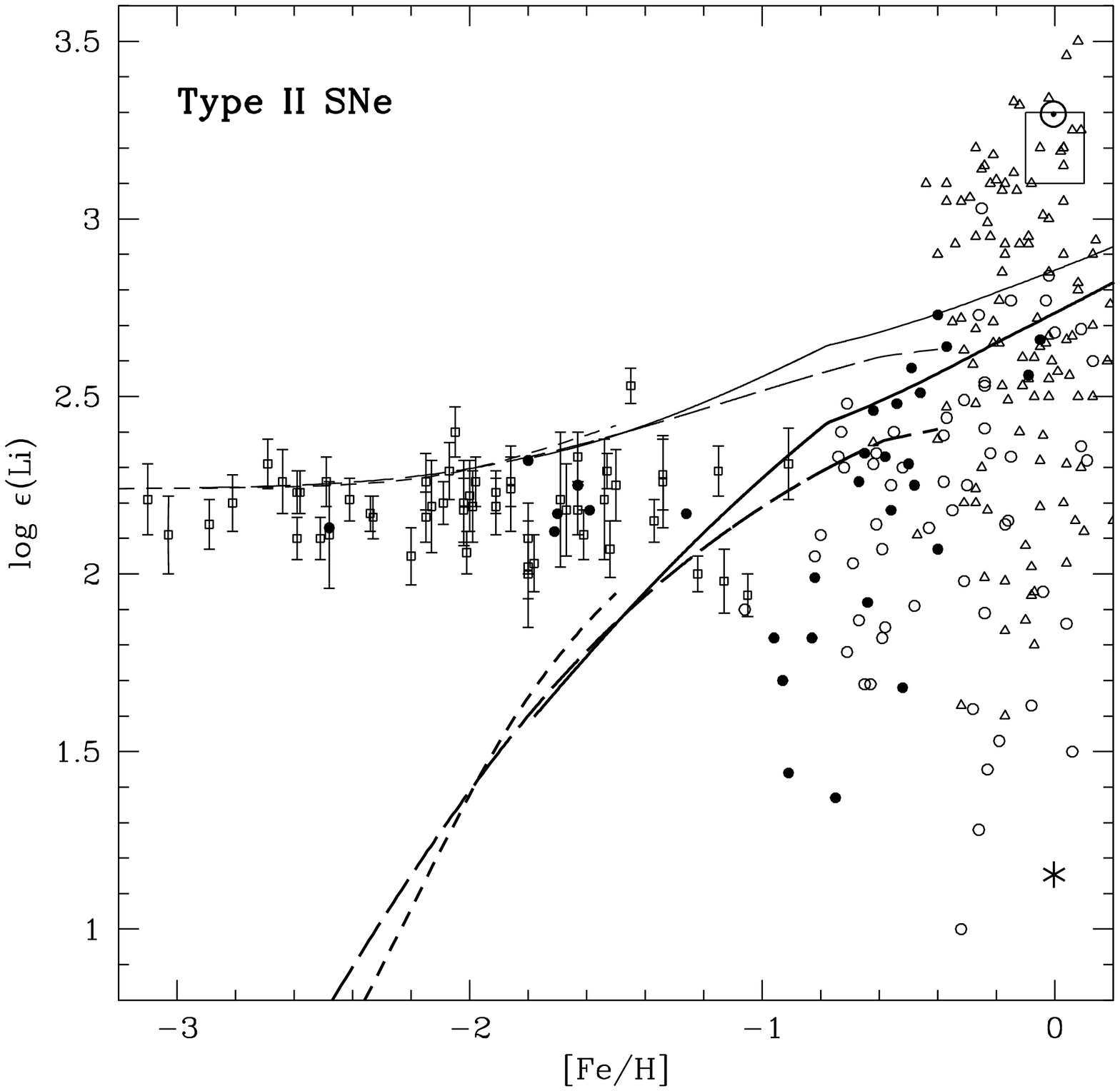}
\figcaption{Contribution of SNII to the Galactic evolution of
\lit (line types are as in Fig.~6). Thick lines
show the \lit evolution obtained with a zero initial abundance of
\lit, to emphasize the contribution of SNII in the thick disk
({\it dashed line}) and in the thin disk ({\it solid line}). Symbols
are as in Fig.1 (upper panel).}
\end{figure}
\clearpage

\begin{figure}
\plotone{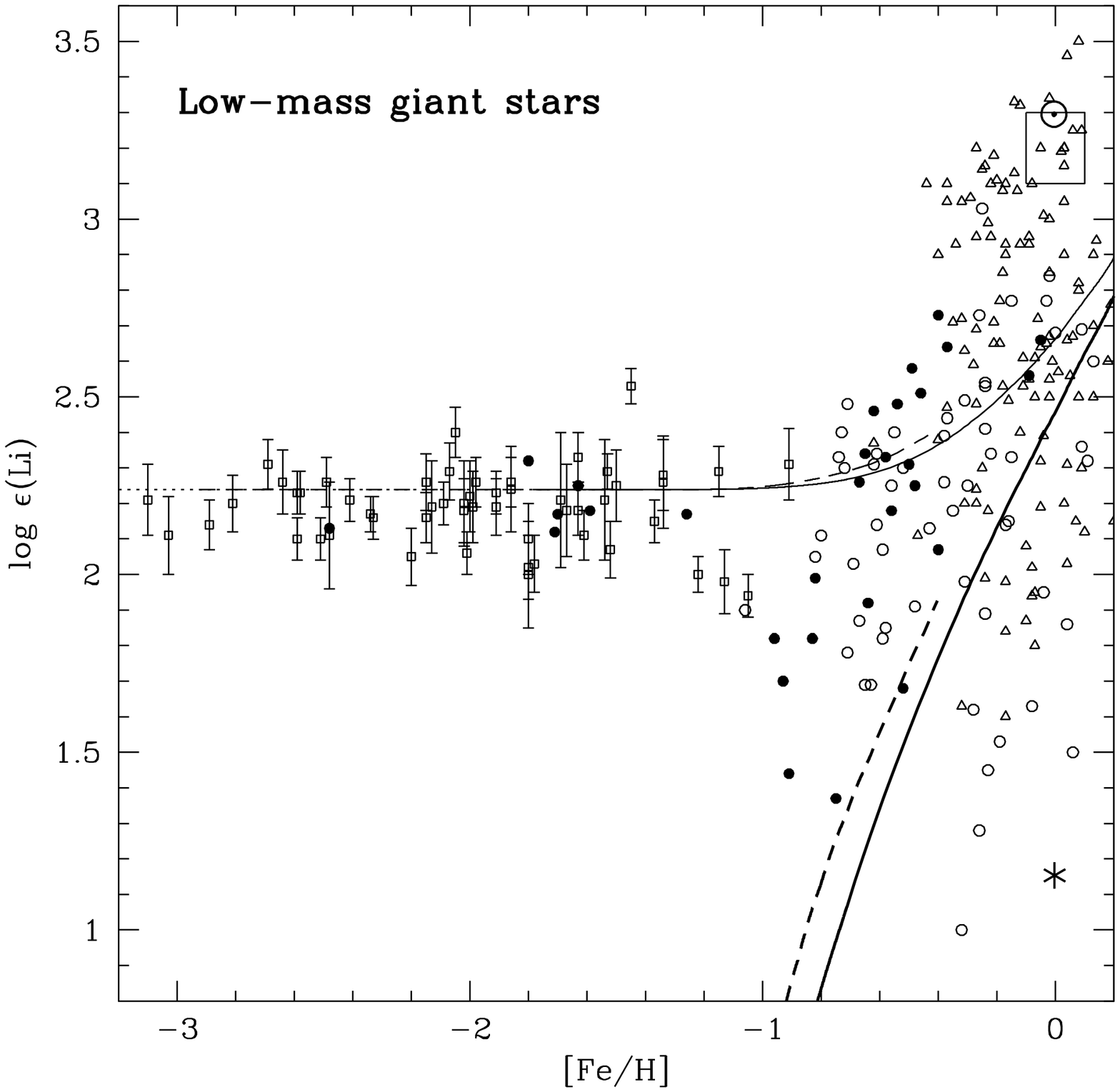}
\figcaption{Contribution of low-mass giant stars via deep mixing
process to the Galactic evolution of \lit (line types are as in Fig.~6).
Thick lines show the \lit evolution obtained
with a zero initial abundance of \lit, to emphasize the contribution of
low-mass giants in the thick disk ({\it dashed line}) and in the thin
disk ({\it solid line}) Symbols are as in Fig.1 (upper panel).}
\end{figure}
\clearpage

\begin{figure}
\plotone{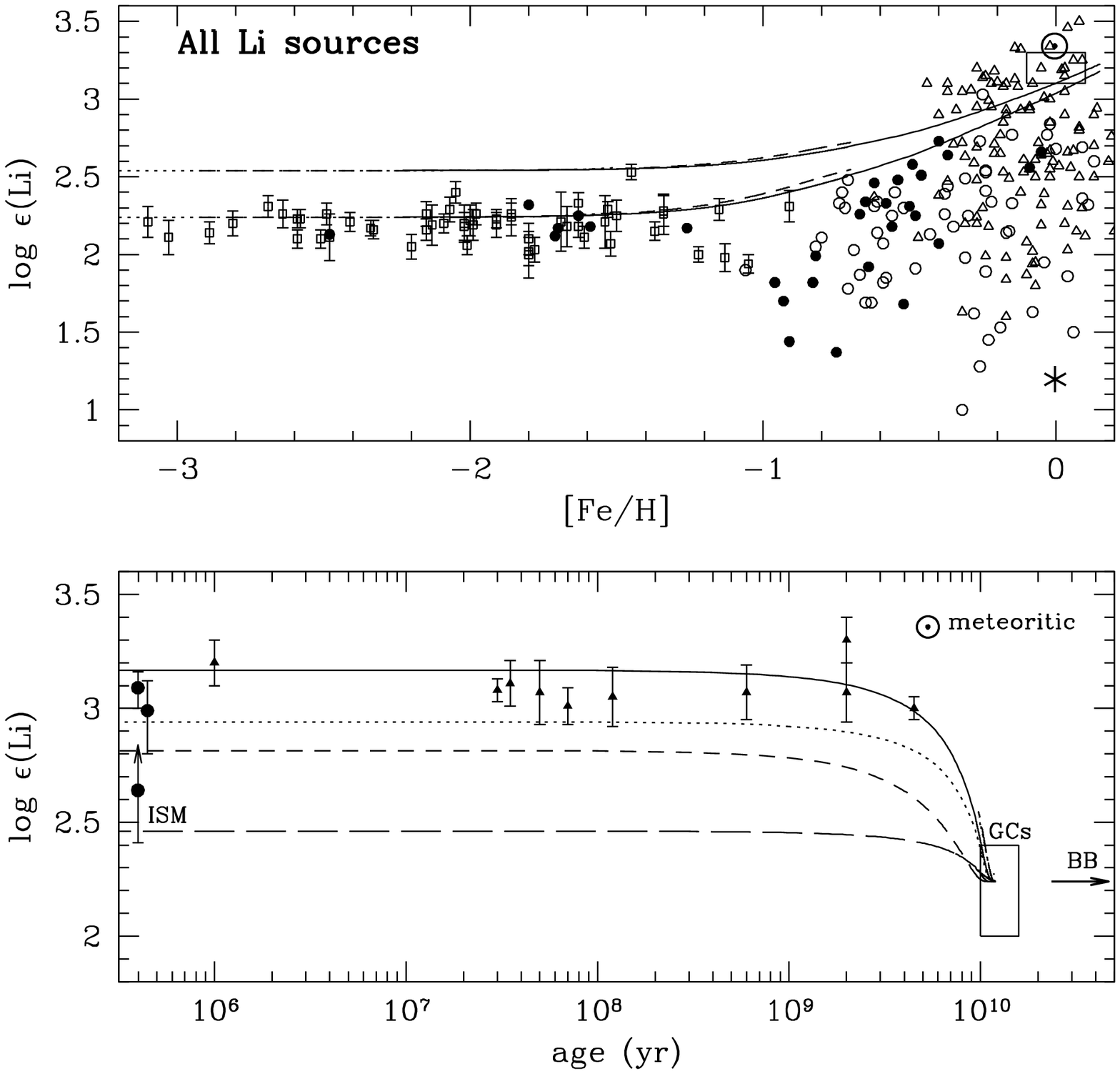}
\figcaption{{\it Upper panel}: evolution of \lit as a
function of [Fe/H] according to our model
taking into account the contribution of novae, low-mass giants and
IMS-AGB stars (line types are as in Fig.~6).  
We also show the results obtained starting with an initial \lit abundance
higher by a factor 2 than the Spite plateau. {\it Lower panel}: 
$\log\epsilon (^7{\rm Li})$ vs. age for Galactic open clusters ({\it filled
triangles}) and globular clusters (same symbols as in Fig.~1).  We
indicate the individual contributions of novae ({\it long-dashed line}),
low-mass giants ({\it short-dashed line}), and IMS-AGB stars ({\it
dotted line}). The {\it solid line} shows the total contribution from all
sources considered.}
\end{figure}


\begin{thebibliography}{}

\bibitem{abir91}Abia, C., Boffin, H.M.J., Isern, J., \& Rebolo, R. 1991, \aap, 245, L1

\bibitem{}Abia, C., Isern, J., \& Canal, R. 1993, A\&A, 275, 96

\bibitem{aic95}Abia, C., Isern, J., \& Canal, R. 1995, \aap, 298, 465

\bibitem{apl99}Abia, C., Pavlenko, Y., \& de Laverny, P. 1999, \aap, 351, 273

\bibitem{ai00}Abia, C., \& Isern, J. 2000, \apj, 536, 438

\bibitem{a67}Alexander, D.R. 1967, The Observatory, 87, 238


\bibitem{ag89}Anders, E., \& Grevesse, N. 1989, Geochim. Cosmochim. Acta, 53, 197

\bibitem{agm01}Arellano Ferro, A., Giridhar, S., Mathias, P. 2001, \aap, 368, 250

\bibitem{an75}Arnould, M., \& N\o rgaard, H. 1975, \aap, 42, 55

\bibitem{b90}Balachandran, S. 1990, \apj, 354, 310

\bibitem{bhfu00}Balachandran, S., Henry, G., Fekel, F.C., \& Uitebroek, H. 2000, 
\apj, 542, 978

\bibitem{b95}Bl\"ocker, T. 1995, \aap, 297, 727

\bibitem{}Bl\"ocker, T., Balega, Y., Hofmann, K.-H., \& Weigelt, G. 2001, 
A\&A, 369, 142

\bibitem{}Boesgaard, A.M., \& Tripicco, M.J. 1986, ApJ, 303, 724

\bibitem{bs85}Boesgaard, A.M., \& Steigman, G. 1985, \araa, 23, 319

\bibitem{bdsk98}Boesgaard, A.M., Deliyannis, C.P., Stephens, A., \& King, J.R.  1998, 
\apj, 493, 206

\bibitem{b93}Boffin, H.M.J., Paulus, G., Arnould, M., \& Mowlawi, N. 1993, \aap, 279, 173

\bibitem{bm97} Bonifacio, P. \& Molaro, P. 1997, \mnras, 285, 847

\bibitem{bs99}Boothroyd, A.I., \& Sackmann, I.J. 1999, \apj, 510, 232

\bibitem{bw91}Bowen, G.H., \& Willson, L.A. 1991, \apj, 375, L53

\bibitem{c55}Cameron, A.G.W. 1955, \apj, 212, 144

\bibitem{cf71}Cameron, A.G.W., \& Fowler, W.A. 1971, \apj, 164, 111

\bibitem{c93}Cannon, R.C. 1993, \mnras, 263, 817

\bibitem{cgcfp00}Carretta, E., Gratton, R.G., Clementini, G., \& Fusi Pecci, F.  2000, 
\apj, 533, 215

\bibitem{cghssb98}Castilho, B.V., Gregorio-Hetem, J., Spite, F., Spite, M., \& Barbuy, 
B.  1998, A\&AS, 127, 139

\bibitem{c94}Charbonnel, C. 1994, \aap, 282, 811


\bibitem{cbw98}Charbonnel, C., Brown, J.A., \& Wallerstein, G. 1998, \aap 332, 204

\bibitem{cb00}Charbonnel, C., \& Balachandran, S.C. 2000, \aap, 359, 563

\bibitem{}Claussen, M.J., Kleinmann, S.G., Joyce, R.R., \& Jura, M. 1987, \apjs, 65, 385

\bibitem{cal95}Coc, A. Mochkovitch, R., Oberto, Y., Thibaud, J.P., \& Vangioni-Flam, E. 
1995, \aap, 299, 479

\bibitem{cr93}Crabtree, D.R., \& Rogers, C. 1993, in Mass Loss on the AGB and beyond, 
ed. H.E. Schwarz (Munich: ESO), p.~255

\bibitem{dm91}D'Antona, F., \& Matteucci, F. 1991, \aap, 247, L37

\bibitem{dal00} de Bernardis, P. et al. 2000, Nature, 404, 955

\bibitem{ds95}de La Reza, R., \& da Silva, L. 1995, \apj, 439, 917

\bibitem{dds96}de La Reza, R., Drake, N.A., \& da Silva, L. 1996, \apjl, 456, L115

\bibitem{ddsm97}de La Reza, R., Drake, N., da Silva, L., \& Mart\'\i n, E. 1997, \apj, 
482, L77

\bibitem{del00}Deliyannis, C.P. 2000, ASP Conference Series 198, p. 235

\bibitem{dr97}Deliyannis, C.P., \& Ryan, S.G. 1997, \apj, 480, L43

\bibitem{del90}Deliyannis, C.P., Demarque, P., and Kawaler, S.D. 1990,
ApJS, 73, 21

\bibitem{del94}Deliyannis, C.P., King, J., Boesgaard, A.M., and
Ryan, S.G. 1994, ApJ, 434, L71

\bibitem{del98}Deliyannis, C.P., Boesgaard, A.M., Stephens, Alex., et al.
1998, ApJ, 498, 147

\bibitem{dv00}Della Valle, M. 2000, in The Chemical Evolution of the Milky Way: stars 
vs. clusters, eds. F. Matteucci \& F. Giovannelli (Dordrecht: Kluwer), p. 371

\bibitem{du91}Duncan, D.K. 1991, \apj, 373, 250

\bibitem{fb93}Fekel, F.C., \& Balachandran, S. 1993, \apj, 403, 708

\bibitem{fg88}Ferrini, F., \& Galli, D. 1988, \aap, 195, 27

\bibitem{fe92}Ferrini, F., Matteucci, F., Pardi, C., \& Penco, U.  1992, \apj, 387, 138

\bibitem{fo99}Fields, B.D., \& Olive, K.A. 1999, \apj, 516, 797

\bibitem{fc97}Forestini, M., \& Charbonnel, C. 1997, \aap, 123, 241

\bibitem{frs70}Fowler, W.A., Reeves, H., \& Silk, J. 1970, \apj, 162, 49

\bibitem{fl96}Frost, C.A., \& Lattanzio, J.C. 1996, \apj, 473, 383

\bibitem{f97}Frost, C.A. 1997, Ph.D. Thesis, Monash University

\bibitem{f98}Frost, C.A., Cannon, R.C., Lattanzio, J.C., Wood, P.R., \& Forestini, M. 1998, \aap, 332, L17

\bibitem{fu00}Fulbright, J.P. 2000, AJ, 120, 1841

\bibitem{gf89}Galli, D., \& Ferrini, F. 1989, \aap, 218, 31  

\bibitem{gal95}Galli, D., Palla, F., Ferrini, F., \& Penco, U. 1995, \apj, 443, 536

\bibitem{gal91}Grevesse, N., Noels, A., \& Sauval, A.J. 1996, in Cosmic Abundances, eds. Holt, S.S. \& Sonnerborn, G., p.~117

\bibitem{gd89}Gratton, R., \& D'Antona, F. 1989, \aap, 215, 66   

\bibitem{hlw00}Heger, A., Langer, N., \& Woosley, S.E. 2000, \apj, 528, 368

\bibitem{hal96}Hernanz, M., Jos\'e, J., Coc, A., \& Isern, J. 1996, \apj, 465, L26 

\bibitem{hp86}Hobbs, L.M., \& Pilachowski, C. 1986, \apj, 309, L17

\bibitem{hd87}Hobbs, L.M., \& Duncan, D.K. 1987, \apj, 317, 796

\bibitem{ib65}Iben, I., 1965, \apj, 142, 1447

\bibitem{ib67a}Iben, I., 1967a, \apj, 147, 624

\bibitem{ib67b}Iben, I., 1967b, \apj, 147, 650

\bibitem{it84}Iben, I., \& Tutukov, A.V. 1984, \apj, 284, 719

\bibitem{jplt99}Jasniewicz, G., Parthasarathy, M., de Laverny, P., \& Th\'evenin, F. 1999, \aap, 342, 831

\bibitem{jef00}Jeffries, R.D. 2000, ASP Conference Ser. 198, p. 245

\bibitem{jh98}Jos\'e, J., \& Hernanz, M. 1998, \apj, 494, 680

\bibitem{kspv99}Klochkova, V.G., Szczerba, R., Panchuk, V.E., \& Volk, K. 1999, \aap, 345, 905

\bibitem{kflc00}Knauth, D.C., Federman, S.R., Lambert, D.L., \& Crane, P. 2000, Nature, 405, 656

\bibitem{lhe91}Lambert, D., Health, J.E., \& Edvardsson, B. 1991, \mnras, 253, 610

\bibitem{lhwh99}Langer, N., Heger, A., Woosley, S.E., \& Herwig, F. 1999, in
Nuclei in the Cosmos V, eds. N. Prantzos \& S. Harissopulos, (Paris: Edition Fronti\`eres), p.~129

\bibitem{lf98}Lattanzio, J.C., \& Forestini, M. 1999, in Asymptotic Giant
Branch Stars, IAU Symposium 191, eds. T. Le Bertre, A. L\`ebre, \& C.
Waelkens, p.~31

\bibitem{lptc}Lattanzio J.C.,  Pettini M., Tout C.A., \& Carigi L,, 2001, \aap, in
press

\bibitem{lhkdm93}Latter, W.B., Hora, J.L., Kelly, D.M., Deutsch, L.K., \& Malony, P.R. 1993, AJ, 106, 1993

\bibitem{llmcs99}L\`ebre, A., de Laverny, P., de Medeiros, J.R., Charbonnel, C., \& da Silva, L. 1999, \aap, 345, L936

\bibitem{lfveb93}Lemoine, M., Ferlet, R., Vidal-Madjar, A., Emerich, C., \& Bertin, P. 1993, \aap, 269, L469

\bibitem{lvfc98}Lemoine, M., Vangioni-Flam, E., Cass\'e, M. 1998, \apj, 499, 735

\bibitem{l00}Lewis, B.M. 2000, \apj, 533, 959

\bibitem{mrp92}Magazz\`u, A., Rebolo, R., \& Pavlenko, Ya.V. 1992, \apj, 392, 159

\bibitem{mrmp94}Mart\'\i n, E.L., Rebolo, R., Magazz\`u, A., \& Pavlenko, Ya.V.  1994, \aap, 282, 503

\bibitem{mm97} Mart\'\i n, E.L., \& Montes, D. 1997, \aap, 318, 805


\bibitem{mdt95}Matteucci, F., D'Antona, F., \& Timmes, F.X. 1995, \aap, 303, 460

\bibitem{mar71}Meneguzzi, M., Audouze, J., \& Reeves, H 1971, \aap, 15, 337

\bibitem{m88}Metzger, P.G. 1988, in Galactic \& Extragalactic Star Formation, 
eds. R. Pudritz, \& M. Fich (Dordrecht: Kluwer), p. 227

\bibitem{nlps99}Nissen, P.E., Lambert, D.L., Primas, F., \& Smith, V.V. 1999, \aap, 348, 211
   
\bibitem{oegc93}Olofsson, H., Eriksson, K., Gustafsson, B., \& Carlstr\"om, U. 1993, \apjs, 87, 267

\bibitem{pbstg00} Palla, F., Bachiller, R., Stanghellini, L., Tosi, M., Galli, D. 2000, \aap, 355, 69

\bibitem{prgc90}Pallavicini, R., Randich, S., Giampapa, M., \& Cutispoto, G. 1990, The Messenger 62, 51
   
\bibitem{pm96}Pasquini, L., \& Molaro, P. 1996, \aap, 307, 761

\bibitem{pm97}Pasquini, L., \& Molaro, P. 1997, \aap, 322, 109

\bibitem{prp97}Pasquini, L., Randich, S., \& Pallavicini, R. 1997, \aap, 325, 535

\bibitem{pb01}Pettini, M., \& Bowen, D. V. 2001, \apj, in press

\bibitem{phs90}Pilachowski, C.A., Hudek, D., \& Sneden, C. 1990, AJ, 99, 1225

\bibitem{p97}Pinsonneault, M.H. 1997, \araa, 35, 557

\bibitem{pcvf93} Prantzos, N., Cass\'e, M., \& Vangioni-Flam, E. 1993, \apj, 403, 630


\bibitem{rslk00}Ramaty, R., Scully, S., Lingenfelter, R., \& Kozlovsky, B. 2000, \apj, 534, 747

\bibitem{rapp97}Randich, S., Aharpour, N., Pallavicini, R., Prosser, C.F., \& 
Stauffer, J.R. 1997, \aap, 323, 86

\bibitem{rmgp98}Randich, S.,  Mart\'\i n, E.L., Garcia Lopez, R.J., \& Pallavicini, R. 1998, \aap, 333, 591

\bibitem{rpp00}Randich, S., Pasquini, L., \& Pallavicini, R. 2000, \aap, 356, L25

\bibitem{ral01}Randich, S., Pallavicini, R., Meola, G., Stauffer, J.R., \& 
Balachandran, S.C. 2001, A\&A, in press

\bibitem{rbm98}Rebolo, R., Beckman, J.E., \& Molaro, P. 1988, \aap, 192, 192

\bibitem{r75}Reimers, D. 1975, in Problems in Stellar Atmospheres and
Envelopes, eds. B. Bascheck, W.H. Kegel, \& G. Traving (Berlin: Springer), p.~229

\bibitem{r81}Renzini, A. 1981, in Physical Processes in Red Giants, eds. I. Jr. Iben, \& A. Renzini (Dordrecht: Reidel), p.~431

\bibitem{rgmpr95}Riera, A., Garc\'ia-Lario, P., Manchado, A., Pottasch, S.R., \& Raga, A.C. 1995, \aap, 302, 137

\bibitem{rmmb99}Romano, D., Matteucci, F., Molaro, P., \& Bonifacio, P. 1999, \aap, 352, 117

\bibitem{rmvd01}Romano, D., Matteucci, F., Ventura, P., \& D'Antona, F. 2001, in Cosmic Evolution, 
eds. M. Cass\'e \& N. Prantzos (Paris: Editions Fronti\`eres), in press

\bibitem{rnb96}Ryan, S.G., Norris, J.E., \& Beers, T.C. 1996, \apj, 471, 254

\bibitem{rnb99}Ryan, S.G., Norris, J.E., \& Beers, T.C. 1999, \apj, 523, 654

\bibitem{r00}Ryan, S.G., Beers, T.C., Olive, K.A., Fields, B.D., \& Norris, J.E. 2000, \apj, 530, L57

\bibitem{rkb01}Ryan, S.G., Kajino, T., Beers, T.C., Suzuki, T.K., Romano, D., Matteucci, 
F., \& Rosolankova, K. 2001, \apj, 249, 55

\bibitem{sb92}Sackmann, I.J., \& Boothroyd, A.I. 1992, \apj, 392, L71

\bibitem{sb99}Sackmann, I.J., \& Boothroyd, A.I. 1999, \apj, 510, 217

\bibitem{sbc99}Salasnich, B., Bressan, A., \& Chiosi, C. 1999, \aap, 342, 131

\bibitem{s87}Schatzman, E. 1987, \aap, 172, 1

\bibitem{sws99}Schr\"oder, K.P., Winters, J.M., \& Sedlmayr, E. 1999, \aap, 349, 898

\bibitem{s97}Shafter, A. 1997, \apj, 487, 226

\bibitem{s80}Shara, M.M. 1980, \apj, 239, 581

\bibitem{sal97}Shara, M.M., Zurek, D.R., Williams, R.E., Prialnik, D., Gilmozzi, R., \& Moffat, A.F.J. 1997, AJ, 114, 258

\bibitem{sl99}Siess, L., \& Livio, M. 1999, \mnras, 308, 1133

\bibitem{sl90}Smith, V.V., \& Lambert, D.L. 1990, \apj, 361, L69

\bibitem{spl95}Smith, V.V., Plez, B., \& Lambert, D.L. 1995, \apj, 441, 735

\bibitem{spfj93}Soderblom, D.R., Pilachowski, C.A., Fedele, S.B., \& Jones, B.  1993, AJ, 105, 2299

\bibitem{sksjf99}Soderblom, D.R., King, J.R., Siess, L., Jones, B.F., \& Fischer, D. 1999, AJ, 188, 1301

\bibitem{ss82}Spite, F., \& Spite, M. 1982, \aap, 115, 357

\bibitem{ss86}Spite, M., \& Spite, F. 1986, \aap, 307, 17

\bibitem{sfns96}Spite, M., Francois, P., Nissen, P.E., \& Spite, F. 1996, \aap, 307, 172

\bibitem{s78}Starrfield, S., Truran, J.W., Sparks, W.M., \& Arnould, M. 1978, \apj, 222, 600

\bibitem{ta73}Talbot, R.J., \& Arnett, D.W. 1973, \apj, 186, 51

\bibitem{tz00} Tegmark, M., \& Zaldiarraga, M. 2000, Phys. Rev. Lett., 85, 2240

\bibitem{th96}Thielemann, F.-K., Nomoto, K., \& Hashimoto, M. 1996, \apj, 460, 408

\bibitem{thdp93}Thornburn, J.A., Hobbs, L.M., Deliyannis, C.P., \& Pinsonneault, M.H.
1993, \apj, 415, 150

\bibitem{tho94} Thorburn, J.A. 1994, \apj, 421, 318

\bibitem{tww95}Timmes, F.X., Woosley, S.E., \& Weaver, T.A. 1995, \apjs, 98, 617

\bibitem{t80}Tinsley, B.M. 1980, Fund. Cosm. Phys., 5, 287

\bibitem{tc86}Tornamb\`e, A., \& Chieffi, A.  1986, \mnras, 220, 529

\bibitem{tggbfs99}Travaglio, C., Galli, D., Gallino, R., Busso, M., Ferrini, F., \& Straniero, O. 1999, \apj, 521, 691

\bibitem{tgbg01}Travaglio, C., Gallino, R., Busso, M., \& Gratton, R. 2001, \apj, in press

\bibitem{vhg89}van der Veen, W.E.C.J., Habing, H.J., \& Geballe, T.R. 1989, \aap, 226, 108

\bibitem{vw93} Vassiliadis, E., \& Wood, P.R. 1993, \apj, 413, 641

\bibitem{vdm00}Ventura, P., D'Antona, F., \& Mazzitelli, I. 2000, \aap, 363, 605

\bibitem{ws82}Wallerstein, G., \& Sneden C. 1982, \apj, 255, 577

\bibitem{wbs95}Wasserburg, G.J., Boothroyd, A.I., \& Sackmann, I.J. 1995, \apj, 447, L37 

\bibitem{w84}Weidemann, V. 1984, \aap, 134, L1

\bibitem{wb72}Wilson, W.J., \& Barrett, A.H. 1972, \aap, 17, 385

\bibitem{w81}Wood, P.R. 1981, \apj, 248, 311

\bibitem{whhh90}Woosley, S.E., Hartmann, D.H., Hoffman, R.D., Haxton, W.C. 1990, \apj, 356, 272

\bibitem{ww95} Woosley, S.E., \& Weaver, T.A. 1995, \apjs, 101, 181
 
\end{thebibliography}
\end{document}